\documentclass[12pt]{iopart}
\pdfoutput=1 
\usepackage[pdftex]{graphicx}
\usepackage{subfigure}
\usepackage{iopams}
\newcommand{\pderiv}[2]{\frac{\partial #1}{\partial #2}}
\newcommand{\ppderiv}[2]{\frac{\partial^{2} #1}{\partial#2^{2}}}
\newcommand{\dint}[1]{\! {\rm d}#1 ~}
\newcommand{\B}[1]{\mathbf{#1}}

\begin{document}

\title[Probability currents in finite populations]{Macroscopically-observable
probability currents in finite populations}

\author{D I Russell and R A Blythe}

\address{SUPA, School of Physics and Astronomy, University of Edinburgh,
Mayfield Road, Edinburgh, EH9 3JZ, UK}

\ead{Dominic.Russell@ed.ac.uk}

\begin{abstract}
In finite-size population models, one can derive Fokker-Planck equations to describe the fluctuations of the species numbers about the deterministic
behaviour. In the steady state of populations comprising two or more species, it is permissible for a probability current to flow. In such a case, the system does not
relax to equilibrium but instead reaches a non-equilibrium steady state. In a two-species model, these currents form cycles (e.g., ellipses) in probability
space. We investigate the conditions under which such currents are solely responsible for macroscopically-observable cycles in population abundances. We
find that this can be achieved when the deterministic limit yields a circular neutrally-stable manifold. We further discuss the efficacy of one-dimensional
approximation to the diffusion on the manifold, and obtain estimates for the macroscopically-observable current around this manifold by appealing to Kramers'
escape rate theory.
\end{abstract}




\maketitle

\section{Introduction}
\label{Secintro}

When deciding how to construct an ecological model to describe the dynamics of a population, a parameter that controls the population size, called the carrying
capacity, $K$, must be taken into account \cite{Trends}. The traditional approach is to assume that $K$ is sufficiently large that fluctuations can be
neglected. Then, a {\it population level model} (PLM) describing the dynamics with ordinary differential equations suffices \cite{Murray}. However for many
biological and ecological systems where populations number typically only hundreds (as opposed to millions, let alone the vast numbers of particles
present in condensed matter systems), it is more appropriate to employ an {\it individual based model} (IBM). Here, the dynamics are cast as stochastic birth
and death processes which can be described as a continuous time Markov process in a discrete configuration space. This configuration space is defined by vector
a $\B{n}=\{n_1,n_2,\ldots,n_N\}$ where $n_i$ is the number of each of the $N$ species in the population, or
equivalently by $\B{x}=\{x_1,x_2,\ldots,x_N\}$  where $x_i=n_i/K$.

The starting point in the analysis of an IBM is the master equation
\begin{equation}
 \pderiv{P(\B{x})}{t}=\sum_\B{x'}[T(\B{x}|\B{x'})P(\B{x')}-T(\B{x'}|\B{x})P(\B{x})] \label{Meqn}
\end{equation}
describing the time evolution of the probability $P(\B{x},t)$ in the space of configurations. The birth and death processes are expressed in terms of the
transition rates $T(\B{x'}|\B{x})$ from configuration $\B{x}$ to $\B{x'}$. By carrying out a systematic system size expansion in $K$,
first prescribed by van Kampen \cite{vanKampen}, of the master equation the deterministic behaviour can be found, recovering the
rate equations of an equivalent PLM description in the $K\to\infty$ limit, and a Fokker-Planck equation (FPE) can be derived to characterise the fluctuations
when $K$ is finite. 

The importance of using the stochastic description of the dynamics when $K$ is finite is revealed by the profound effect noise can have on the behaviour of the
system. A striking example comes from the recent work of McKane and Newman \cite{MckanePredPrey} in the context of a predator-prey system. They show that
although population cycles are absent in the deterministic limit of a simple system, there are sustained oscillations (cycles) in finite populations that can
be explained by  resonant amplification in the demographic noise. The origin of these cycles is a large-amplitude excitation of a deterministic spiral into the
fixed point by the demographic noise in the system. In the language of Tom\'e and de Oliveira \cite{Oliveira}, the demographic noise converts the deterministic,
damped oscillations of the spiral, to undamped oscillations in the form of phase-forgetting quasicycles in the species densities. Subsequently, this phenomenon has been realised in models of systems as diverse as gene regulation
\cite{StoAmpBioChem} and measles epidemics \cite{StoAmpEpi}.

These cycles also appear to be evident in (finite) Lotka-Volterra systems when the stochastic 
dynamics are embedded in real space (see e.g.~\cite{MobGeoTau}). The importance of this work is that it establishes a generic mechanism for sustained population
 cycles in predator-prey systems. The periodic orbits present in the deterministic Lotka-Volterra system, for example, are not robust to modifications 
of the dynamics or noise \cite{Murray}: intrinsic demographic noise has been shown to destabilise marginally stable predator prey cycles \cite{ParkerE, ParkerJ, Abta}, shifting the phase space trajectory
between different limit cycle orbits until eventually one crosses an absorbing state where a species becomes extinct.
 The need for models in which cycles are robustly observed arises from the fact that some natural populations 
indeed exhibit cyclicity in abundances: for example, voles and lemmings exhibit complex multi-year cycles \cite{Stenseth}.
 It is unlikely that such behaviour is achieved by fine-tuning of parameters with highly specific values that happen to favour cyclicity.

In this work, we investigate a potential and hitherto unexplored mechanism for generating cycles in a population dynamical system without parameter tuning.
 It is based on the observation that when one has more than one stochastic variable (as is the case in a multi-species population dynamics)
a \emph{nonequilibrium steady state} that exhibits cyclic probability currents in  configuration space typically arises.  
The question we address in this paper is whether these abstract cycles can be manifested as macroscopically observable cycles in species abundances.
 This is a reasonable hypothesis  since it is known that in \emph{physical}
systems, such currents are directly observable in experiments on optically-tapped colloids \cite{Optic}. There, the presence of a current was
inferred from the persistent bias seen in the evolution of the polar coordinate of the colloid's motion. 

The fundamental origin of these probability currents can be understood from the Fokker-Planck equation \cite{Risken} for the species frequencies $x_i$. This
takes the form
\begin{equation}
\pderiv{P(\mathbf{x},t)}{t}=-\pderiv{}{x_i} J_i(\mathbf{x},t)
\end{equation} 
where the probability current $J_i$ can be expressed via the Kramers-Moyal expansion coefficients $\{\alpha(\mathbf{x}) \}$ as 
\begin{equation}
 J_i=\alpha_iP-\sum_{j}\pderiv{}{x_i}\alpha_{ij}P \;. \label{S}
\end{equation}
One way to obtain a steady state is if currents vanish everywhere, $\B{J}\equiv0$.  This corresponds to an equilibrium steady state in which the
following conditions for detailed balance are met:
\begin{itemize}
 \item The drift term obeys the potential condition: $\partial \alpha_i/\partial x_j=\partial \alpha_j/x_i$\;;
\item The diffusion term is thermal: $\alpha_{ij}=const \times \delta_{ij}$.
\end{itemize}
 In a one-dimensional (1D) system---by which we refer to the size of the phase space; all the work presented here considers non-spatial systems---it is possible to make a change of variable such that these conditions are satisfied,
 unless a current is enforced on the system by virtue of periodic boundary conditions, or driving of the system at the boundaries. If the system has natural boundary conditions, where
the probability and the current vanish at the boundaries, then necessarily the current must be zero everywhere in the steady state. In two or more dimensions, however, detailed balance is
typically only satisfied if it is imposed from the outset, as is appropriate for systems that are at thermal equilibrium.  In particular, when one is dealing
with systems that are defined by their dynamics, rather than by appealing to an energy function, a steady state in which the current is nonzero (but whose
divergence vanishes) is most likely. In two dimensions (2D), the the vanishing of a divergence implies that `streamlines' following the probability current form
closed loops in the 2D phase space within the boundaries of the system, regardless of whether the system has natural, periodic or driven boundary conditions.

The paper is organised as follows. In section \ref{SecPreModel} we examine the simplest 2D model of population dynamics in which a nonequilibrium current can
be induced. This model has a stable fixed point, and the streamlines in probability space are described by ellipses.  However, we show that these
ellipses are not evident macroscopically. The simplest model that we have been able to find that exhibits cyclic behaviour \emph{purely} as a consequence of
nonequilibrium probability currents is set out in section \ref{SecModel}. The deterministic limit of this model has dynamics that has a closed line (or
manifold) of \emph{neutrally-stable} fixed points. The dynamics are such that the system initially evolves towards the manifold. Once it is reached, one does
not expect any further evolution in the deterministic limit: all forces vanish on the manifold. Na\"{\i}vely, one would expect noise to generate unbiased
diffusion around the manifold.  However, the fact that detailed balance is not satisfied leads to a \emph{biased} diffusion around the manifold.  In section
{\ref{SecMeasure} we show that this current can be observed directly in simulations of the stochastic dynamics using the Gillespie algorithm.  We
furthermore show that a good approximation to these dynamics can be obtained using a reduction to a 1D diffusion.  Crucially, the noise in this approximation
is multiplicative which, after transformation to a standard thermal diffusion problem, yields a complex non-conservative forcing.  This dynamics can then be
modelled as a biased diffusion between successive potential minima, the rates of which can be estimated using Kramers' escape rate theory.

\section{Probability currents near a stable fixed point}
\label{SecPreModel}

The simplest population dynamical model with a nonequilibrium steady state has two species, $A$ and $B$  inhabiting a non-spatial  patch of fixed finite size
$N=X_A+X_B+X_E$. Here, $X_A$, $X_B$ and $X_E$ are the numbers of $A$ individuals, $B$ individuals and unoccupied sites.
 Individuals of both species reproduce with rate $a$, while each can die due to: (i) natural death with rate $d$;
 (ii) predation from the other species with rate $p$; and (iii) cannibalism from another of its species with rate $c$.
 Following the formalism for reaction kinetics of \cite{McKanePatches}, these processes give transition rates for the master equation of the form
\begin{eqnarray}
T(X_A+1,X_B|X_A,X_B)&=&2a \frac{X_A}{N}(N-X_A-X_B)  \nonumber\\
T(X_A,X_B+1|X_A,X_B)&=&2a \frac{X_B}{N}(N-X_A-X_B) \nonumber \\
T(X_A-1,X_B|X_A,X_B)&=&dX_A+c\frac{X_A(X_A-1)}{N}+2p\frac{X_AX_B}{N} \nonumber\\
T(X_A,X_B-1|X_A,X_B)&=&dX_B+c\frac{X_B(X_B-1)}{N}+2p\frac{X_AX_B}{N}\;.\label{model1a} 
\end{eqnarray}
The factors of $2$ arise from the combinatorics of the reaction kinetics; the rate constant $d$ has been rescaled by $N$ and $a,p,c$ by $N-1$,
bringing each rate into the van Kampen scaling form, where they are proportional to $N$ \cite{Gardiner,vanKampen}.

We carry out a systematic system size expansion of the master equation \`a la van Kampen \cite{vanKampen} using the ansatz
\begin{equation}
 X_A=N\rho_A(t)+\sqrt{N}x_A(t)\,, \hspace{10mm} X_B=N\rho_B(t)+\sqrt{N}x_B(t) 
\end{equation}
where $\rho_{A,B}$ are well defined densities and $x_{A,B}$ are now the random variables. 
To leading order in $N$, the deterministic behaviour predicts a commonly shared stable fixed point
\begin{equation}
\rho_*=\frac{2a-d}{(2p-c)(4a+2p+c)}\;.
\end{equation}

To the next order in $N$ we find the FPE for the random variables $x_{A,B}$ which describe the fluctuations about this fixed point. Writing the vector
$\B{x}=(x_A,x_B)$, it is
\begin{equation}
 \pderiv{P(\B{x},t)}{t}=-\pderiv{}{x_i}(\gamma_{ij}x_jP)+D_{ij}\frac{\partial^2P}{\partial x_i\partial x_j} \label{sfpFPE}
\end{equation}
where the drift matrix $\gamma$ and the diffusion matrix $D$ 
\begin{eqnarray}
  \gamma&=&-\rho_*\left(\begin{array}{cc}
2a+c & 2(a+p) \label{gammamat} \\
2(a+p) &2a+c \end{array}\right) \\
D&=& 2a\rho_*\frac{c+2(p+d)}{4a+2p+c}\left(\begin{array}{cc}
1 &0 \\
0 &1\end{array}\right)    \label{Dmat}     
\end{eqnarray}
are independent of $\B{x}$. A generic feature of the van Kampen expansion is that this FPE has a drift term
linear in $\B{x}$ and a diffusion matrix that is constant. This implies that its steady-state solution always has the Gaussian form \cite{vanKampen}
\begin{equation}
 P_S(\B{x})=\frac{1}{Z}\e^{-\frac{1}{2}x_iS_{ij}x_j} \label{FPGauss}
\end{equation}
where the inverse covariance matrix $S$ satisfies the constraints
\begin{eqnarray}
 \gamma^TS+S\gamma&=&2SDS\\
 \Tr(DS-\gamma)&=&0 \;. \label{Trace}
\end{eqnarray}

The Gaussian form leads to the steady-state probability current vector $\B{J}$ taking the form
\begin{equation}
 J_i=(DS-\gamma)_{ij}x_jP_S\;. \label{Si}
\end{equation}
Expressing the probability current $\B{J}=\B{\dot x}P_S$ where $\B{\dot x}$ is the average velocity vector and comparing to~(\ref{Si})
we have  for the fluctuations that
\begin{equation}
 \B{\dot x}=(DS-\gamma)\B{x}  \equiv W\B{x}\;. \label{xdotvec}
\end{equation}
For the transitions rates given in~(\ref{model1a}) we find that $W=0$, meaning the system is in thermal equilibrium.
From inspection of (\ref{gammamat}) and (\ref{Dmat}) we see this is because the two necessary conditions for detailed balance are met: the potential condition is satisfied by $\gamma_{12}=\gamma_{21}$ and the diffusion is thermal, $D_{ij}=D_{eq}\delta_{ij}\;$.

We can however induce a nonequilibrium current without modifying the deterministic contribution to the dynamics by introducing the parameters
$r_{1,2}$ and transforming the reaction rates for the birth and death processes for species $A$ (and similarly for $B$) by
\begin{eqnarray}
 a \rightarrow a-r_1\,, \hspace{5mm} d\rightarrow d-2r_1\,, \hspace{5mm}
~p\rightarrow p+r_1\,, \hspace{5mm} ~c\rightarrow c+2r_1\;.
\end{eqnarray}
Then, $\gamma$ remains constant, but the noise is modified so that
\begin{equation}
 D_{ij}=\left[D_{eq}+2r_i\rho_*(2\rho_*-1)\right]\delta_{ij}
\end{equation}
 meaning that if $r_1\neq r_2$ then $D_{11}\neq D_{22}$, making the noise athermal. This is sufficient to render the matrix $W$ in~(\ref{xdotvec})
non-zero, thereby generating a current. The eigenvalues of $W$, which can be expressed as
\begin{equation}
 \lambda_{\pm}=\frac{\Tr{W}}{2}\pm \frac{\sqrt{(\Tr{W})^2-4\det W}}{2}\,, \label{eigenvalues}
\end{equation}
 are purely imaginary due to the constraint~(\ref{Trace}). This means the solution to the autonomous system~(\ref{xdotvec}) are ellipses with a
frequency of orbit 
\begin{equation}
\omega_E=\sqrt{\det W}\;. \label{we}
\end{equation}
We calculate the determinant of $W$ by constructing the Gaussian matrix $S$ in (\ref{FPGauss}) by 
explicitly solving the linear FPE (\ref{sfpFPE}) by the method of characteristics \cite{Risken}. 
\begin{figure}[t]
\begin{center}
\includegraphics[width=0.45\linewidth]{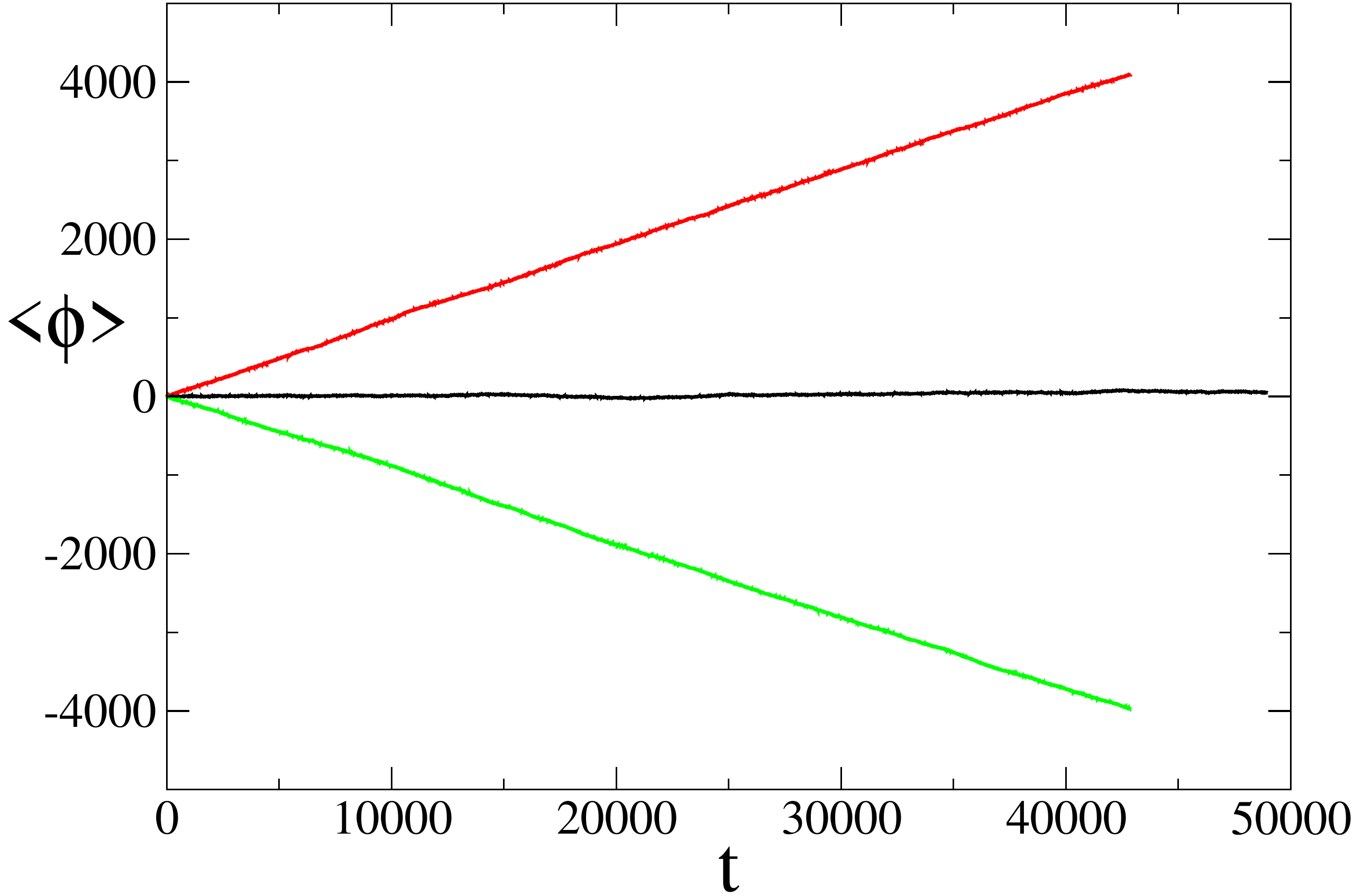}
\caption{Time evolution of the  total average angular displacement $\langle \phi(t)\rangle$ measured by simulation of the stochastic dynamics (\ref{model1a})
by the Gillespie algorithm. The parameters are $a=2.1$, $d=0.5$, $c=2.5$, $p=0.9$. For the red line (top) $r_1=-0.6$, $r_2=0$, the black line (middle)
$r_1=r_2=0$, the green line (bottom) $r_1=0$, $r_2=-0.6$. Each is averaged over $100$ runs.} 
\label{Figbdcgill}
\end{center}
\end{figure}

We now investigate how such a nonequilibrium current is manifested in the macroscopic population dynamics. Our primary tool is simulation of the dynamics
using the Gillespie algorithm \cite{Gillespie}.  To identify a sustained cyclic current in the system, we define for any given transition the angular
displacement $\delta\phi$ about the fixed point $\rho_*$  If we sum up all the $\delta\phi$ over multiple transitions, yielding $\phi(t)$, we should find that
its ensemble average grows linearly in time,  $\langle \phi(t)\rangle \sim\omega_G t$. 

 We see this is borne out in figure \ref{Figbdcgill}, where the criteria for a current flowing are fulfilled through the parameters $r_{1,2}$. For the parameters
quoted in the figure we find by linear regression that the average angular velocity has a magnitude of $\omega_G=0.97$ for the (red) top and (green) bottom plots. 
We also see  that the direction of circulation is determined by the athermality of the noise, i.e. if $D_{11}$ is larger or smaller than $D_{22}$.
This measured speed tallies very well with the frequency of the elliptical orbit calculated by (\ref{we}) for the same parameters: $\omega_E=0.98$.

A crucial feature of the dynamics is that one does not observe cyclicity in population abundances in \emph{single realisations} of the stochastic dynamics on short timescales.  More precisely, it is not possible to see the precession of a single orbit over a $2\pi$ interval. This is because rather than follow a fixed orbital path,
the system diffuses noisily about the fixed point, meaning the cases of the system reaching thermal equilibrium or a nonequilibrium steady state are barely
distinguishable, as is shown in figure \ref{Figsfpnm}.

Therefore we find that a nonequilibrium current could not be advanced as an explanation for cyclic behaviour in natural populations, as observed in single field
experiments.  Moreover, our definition of an angle about the fixed point is tenuous as the system can diffuse very close to and even jump over this origin,
making the notion of an orbit implausible.  We are therefore drawn to the question: Is it possible to set up a population dynamical model so that cyclic
behaviour \emph{is} observed over in a single trajectory over short times?
\begin{figure}[t]
\begin{center}
 \subfigure[\hspace{1mm}$r_1=r_2=0$]{\label{aa}\includegraphics[width=0.32\linewidth]{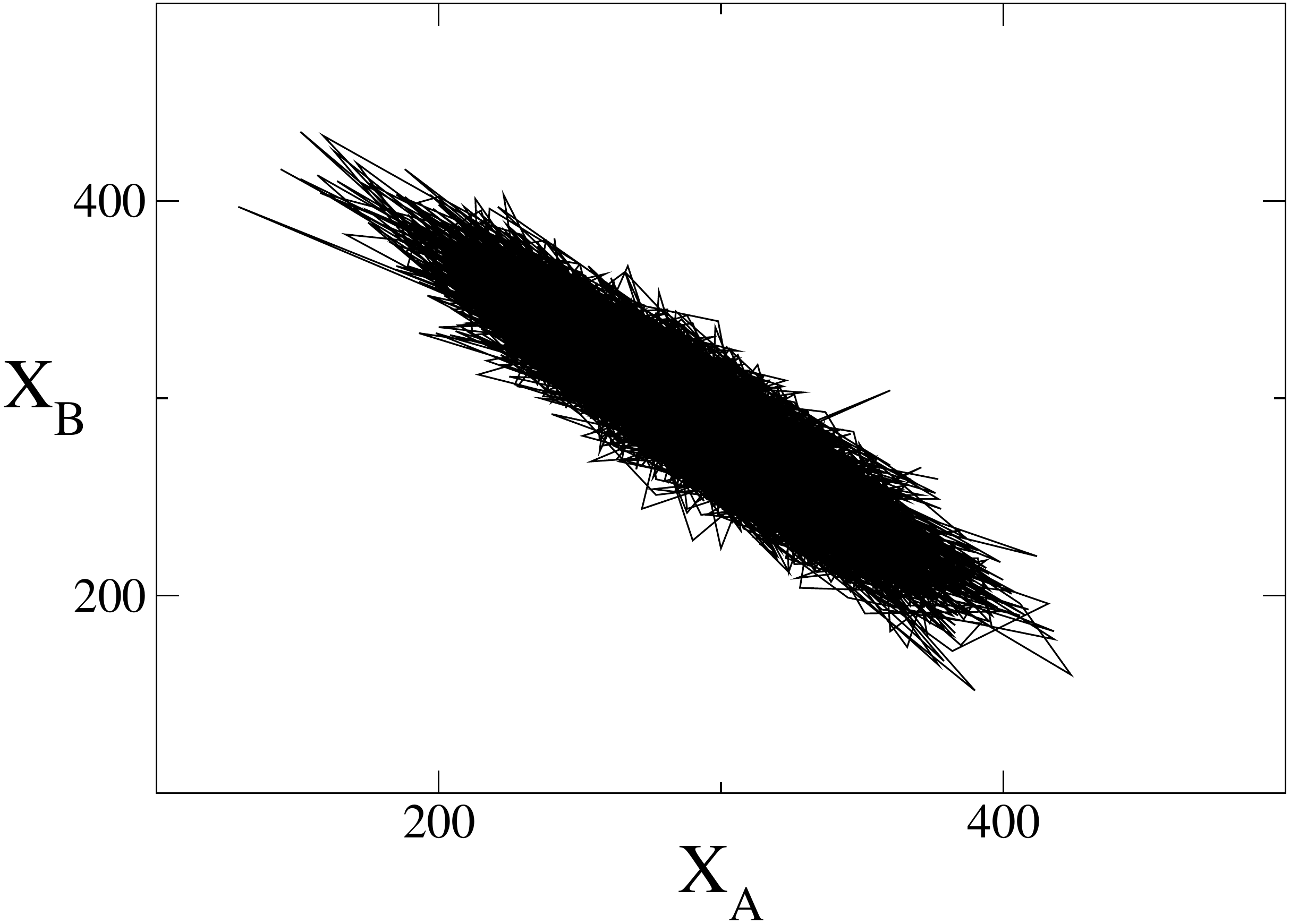}}\hspace{10mm}
 \subfigure[\hspace{1mm}$r_1=-0.6,r_2=0$]{\label{bb}\includegraphics[width=0.32\linewidth]{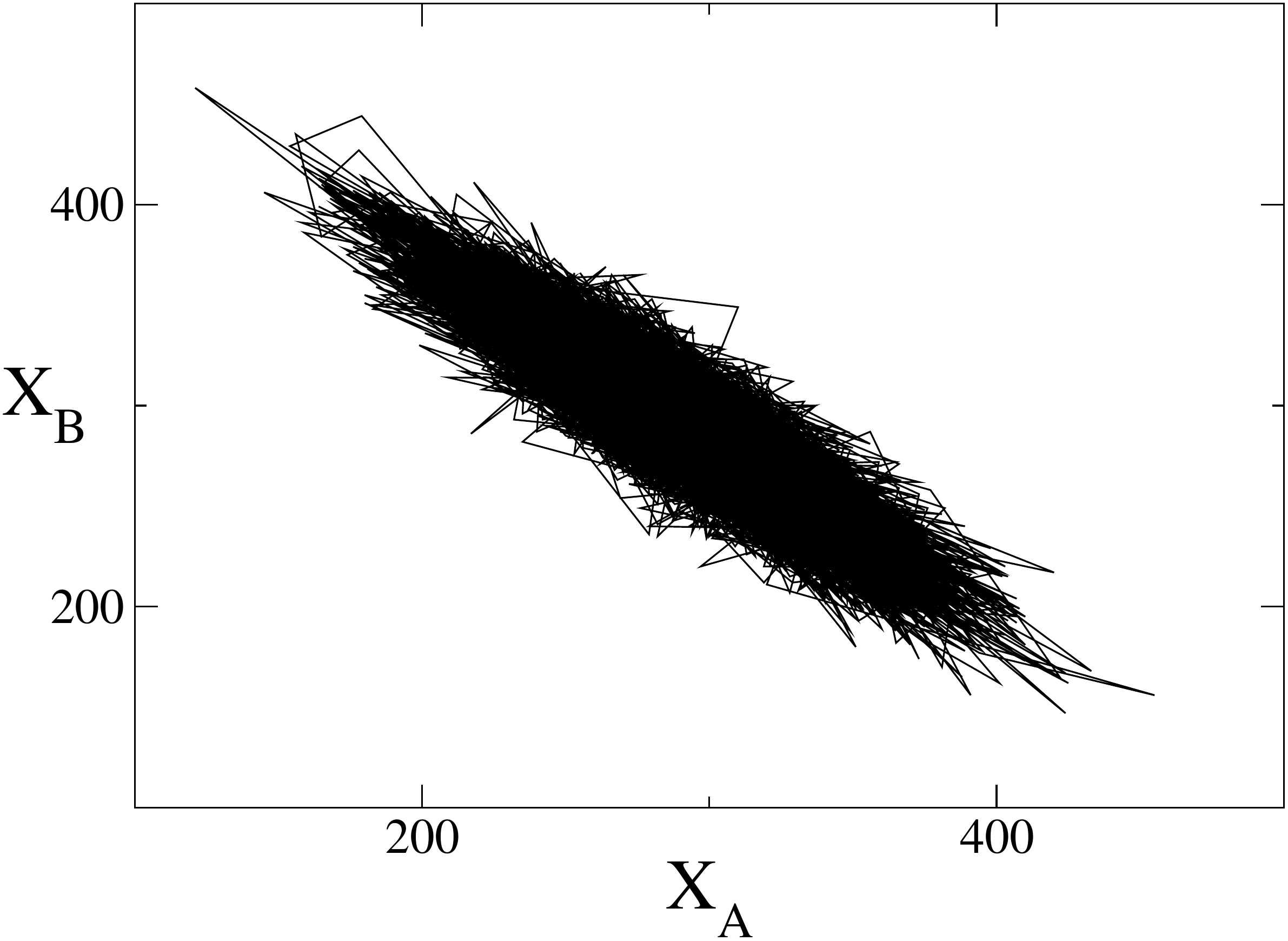}}\hfill
\caption{The phase space of the species $A$ and $B$. The system diffuses noisily about the fixed point, which for
 $a=2.1$, $d=0.5$, $c=2.5$, $p=0.9$ lies at $X_A=X_B=291$.}
\label{Figsfpnm}
\end{center}
\end{figure}

\section{Macroscopic cycles on a neutrally-stable circular manifold}
\label{SecModel}
The basic reason that observable cycles were not evident in the simple model of Section~\ref{SecPreModel} is that the dynamics were attracted deterministically
to a fixed point.  In order for a macroscopic cycle to be evident, what is needed is for the dynamics to be repelled from some origin, so that an orbit may
then be set up.  In particular, if the dynamics are constructed such that the deterministic forcing vanishes along some closed line (or manifold) in phase
space, and such that the dynamics are attracted onto the manifold from outside, then it could in principle be possible for nonequilibrium fluctuations to generate a biased
diffusion on the manifold. If so, in general we would expect to observe cyclical behaviour about the circular manifold.  
We freely admit that, from the perspective of modelling real biological populations, the dynamics that have such a property
are somewhat contrived.  However, from the more fundamental perspective of nonequilibrium statistical mechanics, we believe that this is the simplest such
system that admits macroscopically observable currents arising from nonequilibrium probability currents.  It is also a natural extension of
\emph{quasi-neutral} models of population dynamics from open to closed neutral manifolds.
In \cite{ParsonsQuince}, Parsons and Quince study a model of a non-spatial population with two species, each of which undergo the same birth and death processes.
The model is non-neutral in the sense that the birth and death rate differ for each species. However, when the ratios of these two parameters are the same for
both species, deterministically both population densities evolve to reside at a neutrally stable linear manifold, made up of a continuum of fixed points.
In this sense the model is said to be quasi-neutral. 

This model is defined as follows.It has a similar structure to that of the previous section, except that the size of the non-spatial patch is no longer
conserved, and so the volume of empty space does not explicitly enter into the birth and death rates.  
Defining the intensive population numbers $x_{A}=X_{A}/K,$ $x_{B}=X_{B}/K$ where $K$ is the carrying capacity of the system, both species undergo the following birth and death processes:
\begin{eqnarray}
\fl T(X_A+1,X_B|X_A,X_B)=bx_{A}\left(2+3x^{2}_{A}+x^{2}_{B}+2x_{A}x_{B}\right)\hspace{26mm}\mbox{birth of A} \nonumber\\
\fl T(X_A,X_B+1|X_A,X_B)=bx_{B}\left(2+3x^{2}_{B}+x^{2}_{A}+2x_{A}x_{B}\right)\hspace{26mm}\mbox{birth of B}  \nonumber\\
\fl T(X_A-1,X_B|X_A,X_B)=bx_{A}\left(q+(4-q)x_{A}+2x_{B}+x^{3}_{A}+x_{A}x^{2}_{B}\right)\hspace{5mm}\mbox{death of A} \nonumber\\
\fl T(X_A,X_B-1|X_A,X_B)=bx_{B}\left(q+(4-q)x_{B}+2x_{A}+x^{3}_{B}+x_{B}x^{2}_{A}\right)\hspace{5mm}\mbox{death of B} \label{model} 
\end{eqnarray}
where $b\equiv b(K)$ and $q$ are constants. As will be shown shortly, these rates have been designed to
generate a circular manifold of neutrally-stable fixed points. To interpret them in terms of biological processes, we note that a term in $x_A^{n} x_B^{m}$
corresponds to an interaction between $n$ individuals of species $A$ and $m$ individuals of species $B$. So, for example, the birth of $A$ at a rate
proportional to $x_A^3$ would arise from some interaction between three individuals of species $A$, i.e., some cooperative interaction. Likewise, a
birth rate for $A$ proportional to $x_A x_B^2$ would arise through some interaction between individuals of species $B$: for example, the cooperative
production of some resource by individuals of $B$ that is beneficial to $A$. The higher-order terms in the death rates serve to stop the population sizes running
out of control: they can therefore be interpreted as some kind of resource depletion implied by large populations.  Thus although the combination of
processes that yields a circular neutrally-stable manifold is somewhat specific, the biological principles involved (cooperative behaviour and resource
depletion) are not entirely unreasonable.

For analysis purposes, the useful quantities obtained from the rates are the moments of $\delta x_A$ and $\delta x_B$. These are given by 
\begin{equation}
 \langle (\delta x_A)^i(\delta x_B)^j)\rangle=\sum_{X'_A,X'_B}(\delta x_A)^{i}(\delta x_B)^{j}T(X'_A,X'_B|X_A,X_B)\tau \label{gencartmoms}
\end{equation}
where $X'_{A/B}=X_{A/B}\pm1$ and $T(X'_A,X'_B|X_A,X_B)\tau$ is the probability that $x_A$ and/or $x_B$ changes by $\pm1/K$ in a time $\tau$.
We write the moments in terms of the extensive variables $b$ and $K$, and the intensive variables $x_A$ and $x_B$ via the functions $M_{i,j}(x_A,x_B)$:
\begin{equation}
 \langle (\delta x_A)^i(\delta x_B)^j)\rangle=\frac{b}{K^{i+j}}M_{i,j}\tau\;.\label{Mcartmoms}
\end{equation}
We find for the first and second moments:
\begin{eqnarray}
\fl M_{1,0}=x_{A}\left(2-q-(4-q)x_{A}-2x_{B}+3x^{2}_{A}+x^{2}_{B}+2x_{A}x_{B}-x^{3}_{A}-x_{A}x^{2}_{B}\right) \nonumber \\
\fl M_{0,1}=x_{B}\left(2-q-(4-q)x_{B}-2x_{A}+3x^{2}_{B}+x^{2}_{A}+2x_{A}x_{B}-x^{3}_{B}-x_{B}x^{2}_{A}\right) \nonumber \\
\fl M_{2,0}=x_{A}\left(2+q+(4-q)x_{A}+2x_{B}+3x^{2}_{A}+x^{2}_{B}+2x_{A}x_{B}+x^{3}_{A}+x_{A}x^{2}_{B}\right) \nonumber \\
\fl M_{0,2}=x_{B}\left(2+q+(4-q)x_{B}+2x_{A}+3x^{2}_{B}+x^{2}_{A}+2x_{A}x_{B}+x^{3}_{B}+x_{B}x^{2}_{A}\right) \nonumber \\
\fl M_{1,1}=0\;.\label{M12cartmoms} 
\end{eqnarray}
The deterministic behaviour can be found from the rate equations for $\langle x_A\rangle$ and $\langle x_B \rangle$:
\begin{eqnarray}
\frac{d\langle x_A \rangle}{dt}=\lim_{\tau \rightarrow 0}\frac{\langle x_A(t+\tau)\rangle-\langle x_A(t)\rangle}{\tau}=M_{1,0} \nonumber\\
\frac{d\langle x_B \rangle}{dt}=\lim_{\tau \rightarrow 0}\frac{\langle x_B(t+\tau)\rangle-\langle x_B(t)\rangle}{\tau}=M_{0,1}\,, \label{FPdef}
\end{eqnarray}
can be computed using~(\ref{M12cartmoms}). In the limit $K\rightarrow \infty$ where we neglect fluctuations, $\langle x_A(t) \rangle \equiv x_A(t)$ and 
we find after some algebra:
\begin{eqnarray}
 \dot x_{A}=bx_{A}(x_{A}-1)[q-(x_{A}-1)^{2}-(x_{B}-1)^{2}~] \nonumber\\
 \dot x_{B}=bx_{B}(x_{B}-1)[q-(x_{A}-1)^{2}-(x_{B}-1)^{2}~]\;. \label{fpeqns}
\end{eqnarray}
Defining the polar coordinates $r$ and $\phi$ via
\begin{eqnarray}
x_{A}'\equiv x_{A}-1=r\cos(\phi) \nonumber \\
x_{B}'\equiv x_{B}-1=r\sin(\phi)\,, \label{polarcoord}
\end{eqnarray}
we find that the fixed points of~(\ref{fpeqns}) are
\begin{eqnarray}
(x_{A}=0,x_{B}=0) \label{FPorigin}\\
(x_{A}=1,x_{B}=0) \\
(x_{A}=0,x_{B}=1) \\
(x_{A}=1,x_{B}=1) \label{FPcentre}\\
r=r_{0}=\sqrt{q}\;. \label{FPcircle}
\end{eqnarray}
The fixed point~(\ref{FPorigin}) is an absorbing state at the origin where both species are extinct and is repulsive.
The fixed point~(\ref{FPcentre}) is also repulsive and sits at the centre of the polar coordinate system defined in~(\ref{polarcoord}). 
Physically this  means that $(K,K)$ is the centre of the circle in the extensive system.
Two of the other fixed points are saddle points which sit on the $A$ and $B$ axis respectively, which are absorbing for the respective species.
The fixed point~(\ref{FPcircle}) is a circular manifold of fixed points, located at fixed radius $r_0$ from the centre given by~(\ref{FPcentre}) and
is neutrally stable.

\begin{figure}[t]
\begin{center}
\includegraphics[width=0.45\linewidth]{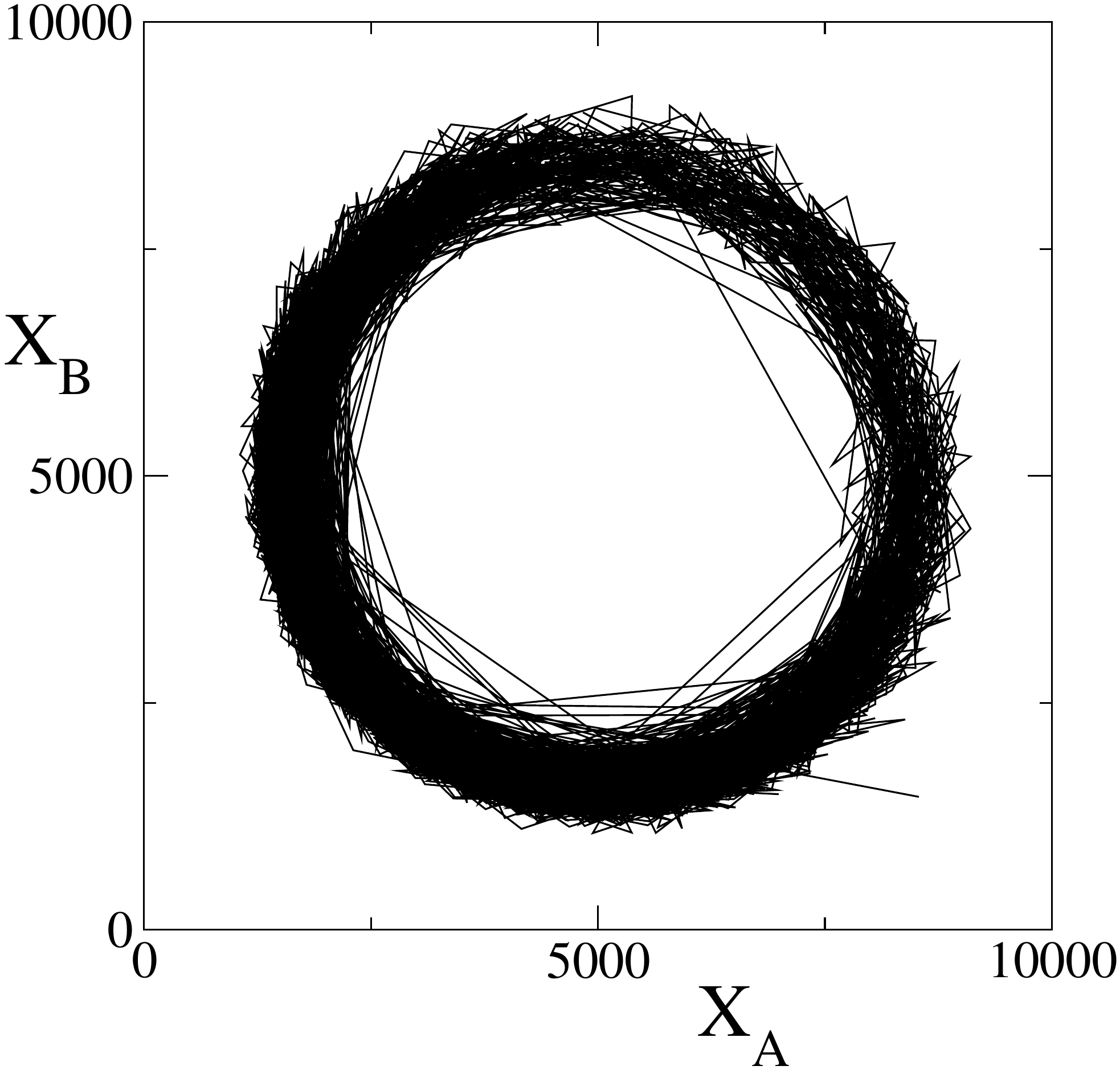}
\caption{Typical dynamical trajectory of the system defined by $X_A(t)$ and $X_B(t)$ and the transition rates given in (\ref{model}).
These data were obtained by simulation using the Gillespie algorithm of the dynamics with parameters $K=5000$, $b=K^2$, $q=0.5$.} 
\label{Figtypstate}
\end{center}
\end{figure}

To summarise,  we find that at the level of the deterministic equations, the system evolves to a fixed point on the circular manifold. The fixed point that is reached would be determined by the initial conditions.
For finite $K$ however, we expect the system to  evolve to the neighbourhood of the manifold, and then diffuse about it. A typical snapshot of the evolution 
of the system is given in figure~\ref{Figtypstate}: this indeed shows that the occupied region of phase space is a 
circular annulus of large radius relative to its width.

\section{Reduction to a one-dimensional diffusion}
\label{SecQuasi1D}

To analyse the model introduced in the previous section, we reduce the full dynamics in the two-dimensional space spanned by $x_A$ and $x_B$ to the
diffusion of the angular coordinate $\phi$ in a polar coordinate system defined in (\ref{polarcoord}) whose origin is the centre of the circle.
A similar approach was used by Parker and Kamenev in studying the stability of a stochastic Lotka-Volterra model \cite{ParkerJ}.
There, as the angular component of the motion relaxes rapidly, it can be integrated out of the probability distribution, allowing the dynamics of the 2D system
to be described by the 1D stochastic radial motion between the deterministic limit cycles. 
In contrast, our main assumption is to neglect diffusion in the radial $r$ direction off the stable circular manifold. That is, we assume that the restoring force 
that acts perpendicular to the manifold is sufficiently strong that any deviation away from the manifold does not contribute to the dynamics in an important way.
We note, however, that this lateral diffusion has been seen to enter into an effective description on an open manifold \cite{ParsonsQuince}. Here, our aim is
to see how well we can understand the full two-dimensional diffusion, and in particular any sustained angular velocity, within a highly simplified
approximation. In a very recent piece of work \cite{nullcline}, Constable {\it et al} studied the dynamics of a two-species population which deterministically
evolves to exist on a stable hyperbolic manifold. They carried out a similar dimensional reduction to that presented here and found it to be
a good approximation of the full 2D dynamics. However a crucial difference is that the hyperbola has natural boundary conditions. As mentioned previously
for a 1D system, natural boundary conditions means the current must vanish at the boundaries. 
Therefore the system cannot support a sustained constant current in the steady state as it must be zero everywhere on the hyperbola.

In order to study the diffusion in the $\phi$ direction around the manifold, we derive a polar FPE for the evolution of the probability density
$P(\phi,t)$, details of which can be found in~{\ref{AppFP}. It is
\begin{equation}
 \pderiv{P(\phi,t)}{t}=-\pderiv{}{\phi}\bigg[f(\phi)P\bigg]+\frac{1}{2}\ppderiv{
}{\phi}\bigg[g^{2}(\phi)P\bigg] \label{FPphi}
\end{equation}
where the drift term $f$ and the diffusion term $g$ are
\begin{eqnarray}
\fl f(\phi)=\frac{1}{4r_0^2}\bigg(2r_0(12+r_0^2)[\sin(\phi)-\cos(\phi)]+4r_0(6+r_0^2)[\sin(3\phi)+\cos(3\phi)] \nonumber \\
+20r_0^2\sin(4\phi)+2r_0^3[\sin(5\phi)-\cos(5\phi)]\bigg)\,, \label{f}
\end{eqnarray}
\begin{eqnarray}
\fl g^2(\phi)=\frac{1}{4r_0^2}\bigg(64+28r_0^2+r_0(56+6r_0^2)[\sin(\phi)+\cos(\phi)]+4r_0(6+r_0^2)[\sin(3\phi)-\cos(3\phi)]& \nonumber\\
+24r_0^2\sin(2\phi)-20r_0^2\cos(4\phi)-4r_0^2[\sin(5\phi)-\cos(5\phi)]\bigg)\;.& \label{g}
\end{eqnarray}
We can write the equivalent Langevin equation
\begin{equation}
 \dot \phi=f(\phi)+g(\phi)~\eta_{\phi} \label{LEphi}
\end{equation}
using the It\^o prescription \cite{Risken} where $\eta_\phi$ is Gaussian white noise with zero mean and unit variance.
As one would expect, $f$ and $g$ are $2\pi$-periodic as can be seen in figure~(\ref{Figfg}).
\begin{figure}[t]
\begin{center}
 \subfigure[\hspace{1mm}$f(\phi)$]{\label{a}\includegraphics[width=0.26\linewidth]{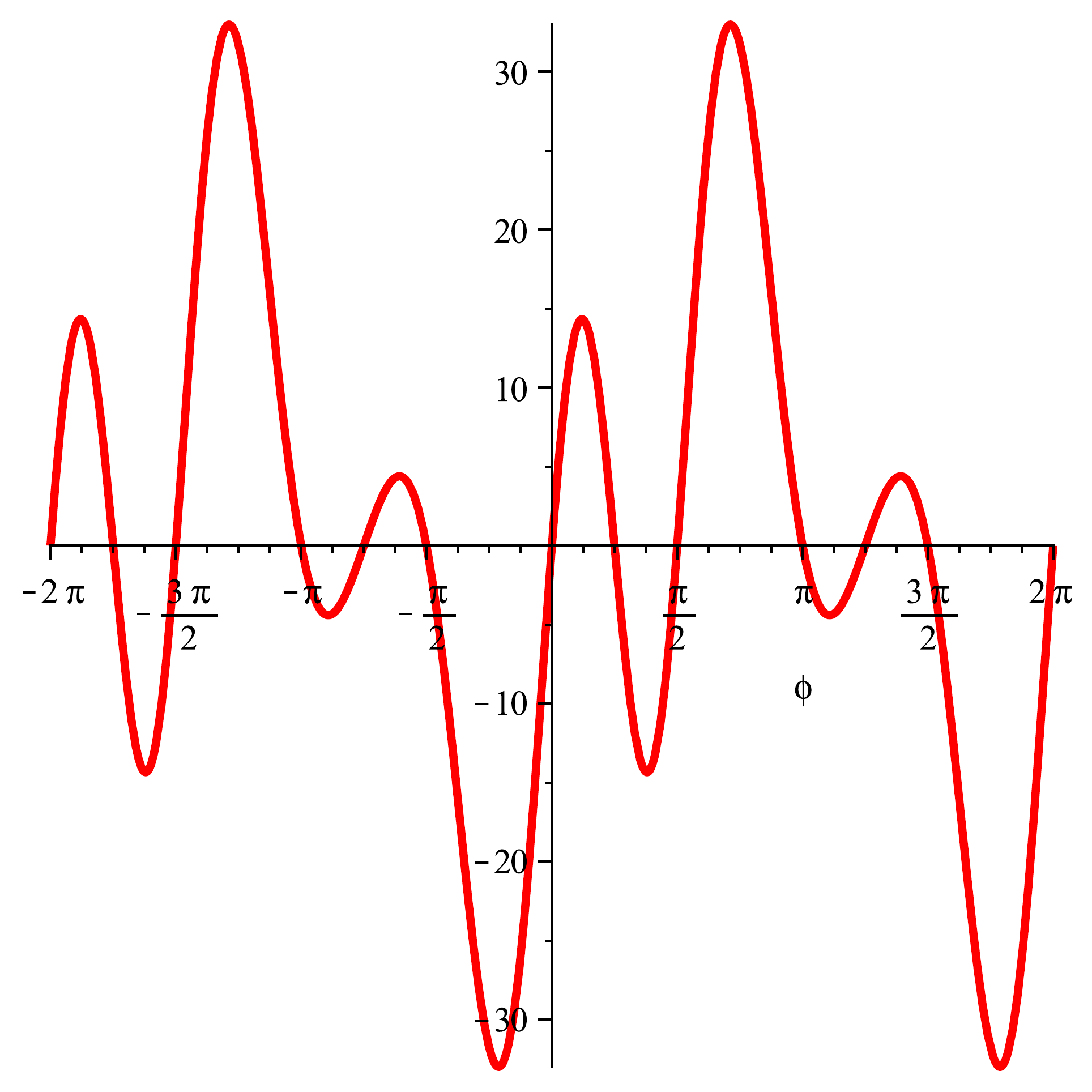}}\hspace{10mm}
 \subfigure[\hspace{1mm}$g(\phi)$]{\label{b}\includegraphics[width=0.26\linewidth]{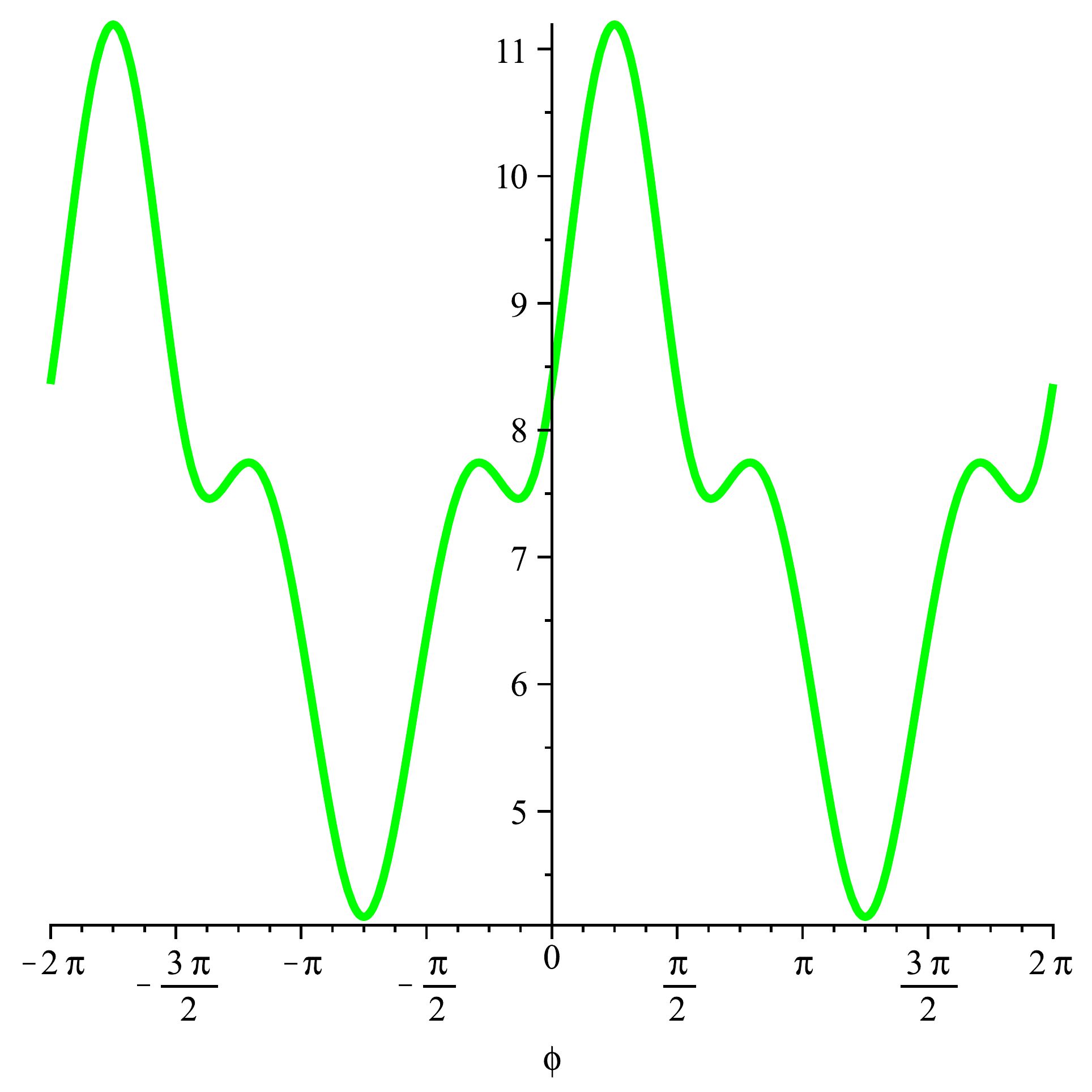}}\hfill
\caption{The drift term $f$ and the diffusion term $g$ of the Langevin equation (\ref{LEphi}) for $q=0.5$.}
\label{Figfg}
\end{center}
\end{figure}

A key feature of the Langevin equation~(\ref{LEphi}) is that the noise is multiplicative. The criteria for a nonequilibrium steady state
to exist in a system with a FPE of the form (\ref{FPphi}) with periodic boundary conditions are more easily checked in the case where the noise is additive.
It is always possible to perform a change of variable which transforms a 1D FPE with athermal diffusion to one with thermal diffusion \cite{Risken}. 
The required change of variable is
\begin{equation}
 \theta(\phi)=\sqrt{2D}\int_{0}^{\phi}\frac{d\phi'}{g(\phi')}\,,\label{phitothetadef}
\end{equation}
under which the FPE~(\ref{FPphi}) becomes
\begin{equation}
 \partial_{t}
Q(\theta,t)=-\partial_{\theta}\left(F(\theta)Q\right)+{D}\partial^{2}_{\theta}Q
\label{FPtheta}
\end{equation}
describing the time evolution of the probability density $Q(\theta,t)$ where now the diffusion is a constant $D$ and the drift force $F(\theta)$ is
\begin{equation}
 F(\theta)=\sqrt{2D}~\left(\frac{f(\phi)}{g(\phi)}-\frac{1}{2}\partial_{\phi}g(\phi)\right)\bigg|_{\phi=\phi(\theta)} \;. \label{Ftheta}
\end{equation}
The corresponding Langevin equation (\`a la It\^o) is
\begin{equation}
 \dot \theta=F(\theta)+\sqrt{2D}~\eta_{\theta} \;.\label{thetalangevin}
\end{equation}
In this transformation one can choose $D$ arbitrarily. For completeness we will proceed with arbitrary $D$, but in numerical calculations will set it to unity.

Unfortunately, it is difficult to obtain an analytic expression for $F(\theta)$, as one needs to evaluate the integral (\ref{phitothetadef}) and invert the
resulting expression to obtain the function $\phi(\theta)$ that appears in (\ref{Ftheta}).  It is however possible to determine whether the drift force
$F(\theta)$ is conservative, and hence infer whether the system will reach thermal equilibrium, without an explicit expression for this function.  

First, we note that $F(\theta)$ is conservative if the work done in one circulation of the manifold vanishes, i.e., if
\begin{equation}
 \oint \dint{\theta}F(\theta)=0 \;.
\end{equation}
We can write an ansatz for $F$:
\begin{equation}
 F(\theta)=\omega-\partial_{\theta}V(\theta)\,, \label{Fansatz}
\end{equation}
where the first term $\omega$ is a constant, non-equilibrium driving force and the second term is a conservative force derived from a potential $V(\theta)$. 
To determine whether $\omega=0$, formally, we should be able to write $F(\theta)$ as a Fourier series
\begin{equation}
F(\theta) = \frac{1}{2} a_0 + \sum_{n=1}^{\infty} \left[ a_n \cos \left( \frac{n\theta}{T} \right) + b_n \sin \left( \frac{n \theta}{T} \right) \right]\label{fourierseries}
\end{equation}
where, using (\ref{phitothetadef}),
\begin{equation}
 T = \frac{\sqrt{2D}}{2} \int_{-\pi}^{\pi} \frac{{\rm d} \phi}{g(\phi)} \label{Tint}
\end{equation}
is the half-width of the transformed interval $[-\pi,\pi]$, viz, $T=\theta(\pi)-\theta(0)$. (We can take $\theta(0)=0$ with no loss of generality).
The coefficients are given in the usual way as
\begin{eqnarray}
a_n &=& \frac{1}{T} \int_{-T}^{T}\! {\rm d} \theta\, F(\theta) \cos\left(\frac{n \theta}{T} \right) \\
b_n &=& \frac{1}{T} \int_{-T}^{T}\! {\rm d} \theta\, F(\theta) \sin\left(\frac{n \theta}{T} \right) \;.
\end{eqnarray}
Comparing~(\ref{Fansatz}) and~(\ref{fourierseries}) we see that $\omega= a_0/2$. So the question of whether $F(\theta)$ is conservative and hence whether
the system is reaches thermal equilibrium is equivalent to finding if $a_0$ is zero.

Taking
\begin{equation}
a_0 = \frac{1}{T} \int_{-T}^{T}\! {\rm d} \theta\, F(\theta)
\end{equation}
we apply the change of variable (\ref{phitothetadef}):
\begin{equation}
a_0 = \frac{1}{T} \int_{\phi(-T)}^{\phi(T)} \! {\rm d} \phi\, F(\theta[\phi])\frac{{\rm d} \theta}{{\rm d} \phi} \;.\label{a0phitotheta1}
\end{equation}
By definition $\phi(\pm T) = \pm\pi$, $F(\theta[\phi])$ can be obtained directly by~(\ref{Ftheta}), formally via $\phi(\theta[\phi]) = \phi$, and we know the
Jacobian of the transformation. So,
\begin{eqnarray}
a_0 &=& \frac{\sqrt{2D}}{T} \int_{-\pi}^{\pi}  \! {\rm d} \phi\,\left(\frac{f(\phi)}{g(\phi)^2} - \frac{{\rm d}}{{\rm d} \phi} \ln g(\phi) \right) \label{a0phitotheta2} \\
&=& \frac{\sqrt{2D}}{T} \int_{-\pi}^{\pi}  \! {\rm d} \phi\,\frac{f(\phi)}{g(\phi)^2} \label{a0int} \,, 
\end{eqnarray}
since the second term vanishes due to the periodicity of $g(\phi)$.  

Although we have successfully side-stepped the problem of evaluating $\phi(\theta)$, it remains the case that the integral~(\ref{a0int}) does not
have a convenient closed form. However we can determine whether it is zero by finding if the integrand
\begin{equation}
 h(\phi)=\frac{f(\phi)}{g^2(\phi)}
\end{equation}
is odd over the limits of integration. In figure~\ref{Figfg} we see that $f$ is odd and $g$ is even about $\phi=\pi/4$. This implies that $h$ is also
odd about $\pi/4$, namely:
\begin{equation}
 h(\phi)=-h\bigg(\frac{\pi}{2}-\phi\bigg)\;. \label{hodd}
\end{equation}
Due to the $2\pi$-periodicity of $h$ we can write (\ref{a0int}) as
\begin{equation}
a_0=\int_{-\pi}^{\pi}\dint{\phi} h(\phi)=\int_{-\frac{3\pi}{4}}^{\frac{5\pi}{4}}\dint{\phi} h(\phi)
 =\int_{\frac{\pi}{4}}^{\frac{5\pi}{4}} \dint{\phi}h(\phi)+\int_{-\frac{3\pi}{4}}^{\frac{\pi}{4}} \dint{\phi} h(\phi)\;.
\end{equation}
Changing variable to $\beta=\pi/2-\phi$ in the second integral we have
\begin{equation}
a_0=\int_{\frac{\pi}{4}}^{\frac{5\pi}{4}} \dint{\phi}h(\phi)+\int_{\frac{\pi}{4}}^{\frac{5\pi}{4}}  \dint{\beta}h\big(\frac{\pi}{2}-\beta \big)\;.
\end{equation}
Now applying (\ref{hodd}) we see that $a_0=0$.

Intuitively it seems correct that the system reaches thermal equilibrium as the model defined by the processes in (\ref{model}) 
is neutral and each species undergoes the same processes at equivalent rates. The insight that is gained from this analysis is that we now know in order
to reach a nonequilibrium steady state, we must introduce  processes which will enter the drift term $f$ and the diffusion $g$ in such a way
as to make the integrand in (\ref{a0int}) not be odd. As in Section~\ref{SecPreModel}, we are interested in the effect of different forms of the noise 
whilst holding the deterministic contribution to the dynamics constant. To this end we introduce the following extra rates to our model:
\begin{eqnarray}
 T(X_A+1,X_B|X_A,X_B)=b(p_1x_{A}x_B^3+p_2x_a^3x_B) \nonumber\\
 T(X_A,X_B+1|X_A,X_B)=b(p_1x_{A}x_B^3+p_2x_a^3x_B)   \nonumber\\
 T(X_A-1,X_B|X_A,X_B)=b(p_1x_{A}x_B^3+p_2x_a^3x_B) \nonumber\\
 T(X_A,X_B-1|X_A,X_B)=b(p_1x_{A}x_B^3+p_2x_a^3x_B) \;. \label{newprocesses} 
\end{eqnarray}
These new rates cancel in the first moments $\langle x_A \rangle$ and $\langle x_B \rangle$ and so leave the deterministic equations~(\ref{FPdef}) unaltered.
The second moments accrue the extra terms
\begin{eqnarray}
 \langle (\delta x_{A})^2\rangle&=\frac{b}{N_0^2}\bigg[2p_1x_Ax_B^3+2p_2x_A^3x_B\bigg]\tau\\
\langle (\delta x_{B})^2\rangle&=\frac{b}{N_0^2}\bigg[2p_1x_Ax_B^3+2p_2x_A^3x_B\bigg]\tau \;.
\end{eqnarray}
Substituting these into~(\ref{f}) leaves $f$ as it was. However the form of $g$ will change.
The two new parameters $p_1$ and $p_2$ will determine the magnitude and direction of the probability current. The new terms that appear in our expression
for $g$ (\ref{g}) are 
\begin{equation}
 \fl 2p_1\sin^2(\phi)(1+r\cos(\phi))(1+r\sin(\phi))^3 +2p_2\cos^2(\phi)(1+r\sin(\phi))(1+r\cos(\phi))^3\;. \label{newgterms}
\end{equation}
From this we see that $g$ will only remain even about $\pi/4$ as long as $p_1=p_2$. 
Therefore as long as $p_1\neq p_2$ a probability current will flow in the steady state. This condition breaks the neutral selection of the model
as the rates of birth and death for each species are no longer exactly equivalent. However \emph{quasi-neutrality} is maintained in the sense that the deterministic
behaviour of both species is the same \cite{ParsonsQuince}, each evolving to reside on the common circular manifold.

\section{Methods for measuring the current}
\label{SecMeasure}

We now outline three approaches to measuring the current in the nonequilibrium steady state.  The first is to use Monte Carlo simulation of the dynamics for
the full problem. The second is to numerically integrate the Langevin equation within the one-dimensional reduction, thereby revealing any error that is
introduced in this procedure. The third is to appeal to Kramers' escape-rate theory to estimate the current.  Our expectation is these methods trade accuracy
for precision and (in the latter case) analytical insight.

\subsection{Monte Carlo simulation}

The probability current can be written as
\begin{equation}
 J(\phi,t)=P(\phi,t) \omega \label{j}
\end{equation}
where $P$ is the probability density and $\omega$ is the average angular velocity. As already mentioned, we infer the flow of a current from a non-zero $\omega$. 
Using the Gillespie algorithm the stochastic model as defined by the rates in~(\ref{model}) and~(\ref{newprocesses}) can be simulated. 
We measure the total angular displacement $\phi_G(t)$ which is positive in the anti-clockwise direction and quantifies the total distance travelled.
 For each update
\begin{equation}
 \phi_G(t+\tau)=\phi_G+\delta \phi
\end{equation}
where $\delta \phi$ is calculated using~(\ref{deltaphi}). In figure~\ref{Figtypgillphi} we plot the angular displacement and find that 
\begin{equation}
 \langle \phi_G(t) \rangle \sim \omega_G~t \;.
\end{equation}
For the parameters quoted, linear regression of the data yields $\omega_G=-0.97$.
\begin{figure}[t]
\begin{center}
 \subfigure[\hspace{1mm}Simulation]{\label{Figtypgillphi}\includegraphics[width=0.32\linewidth]{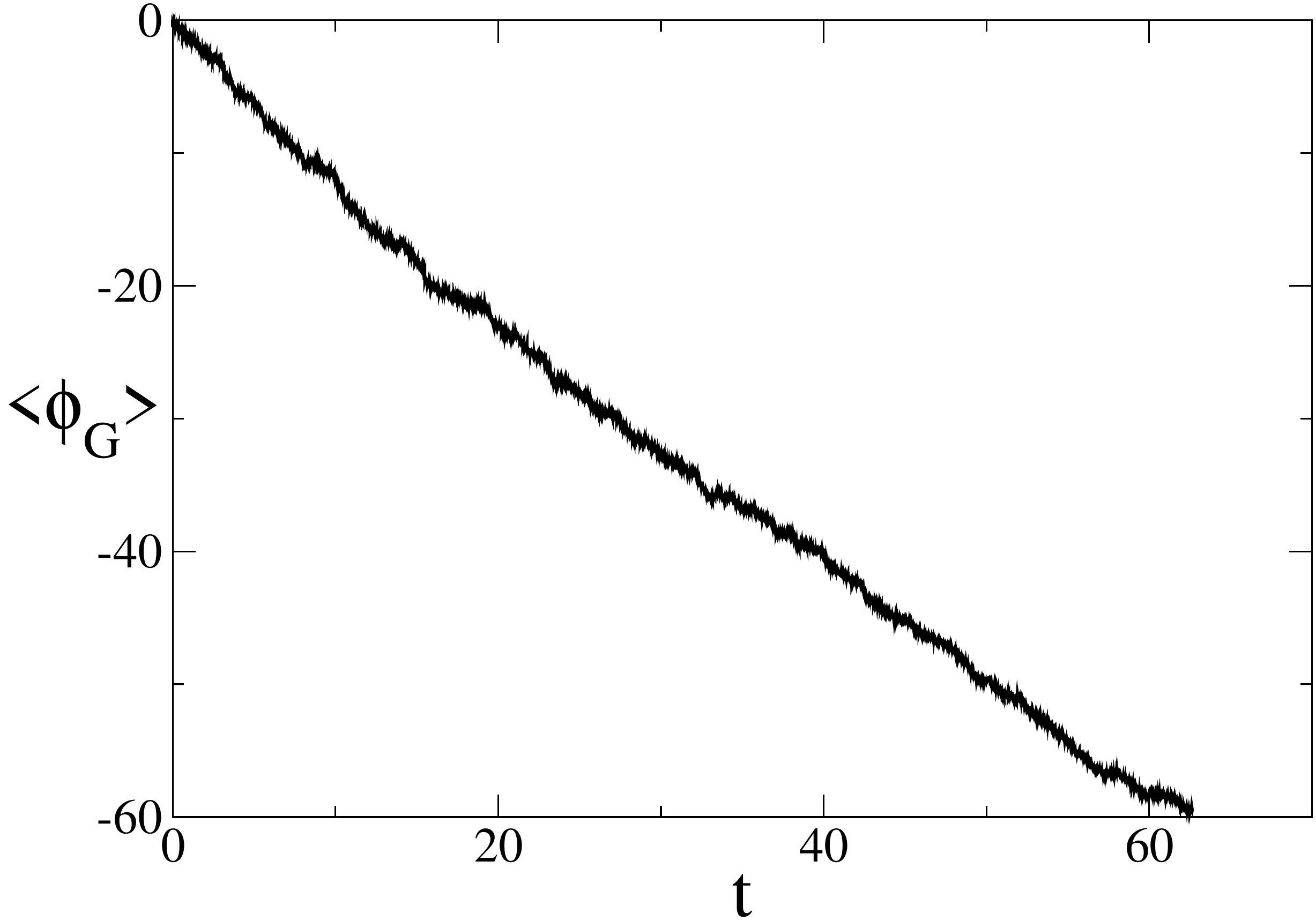}}\hspace{10mm}
  \subfigure[\hspace{1mm}Numerical Integration]{\label{Figtyplephi}\includegraphics[width=0.32\linewidth]{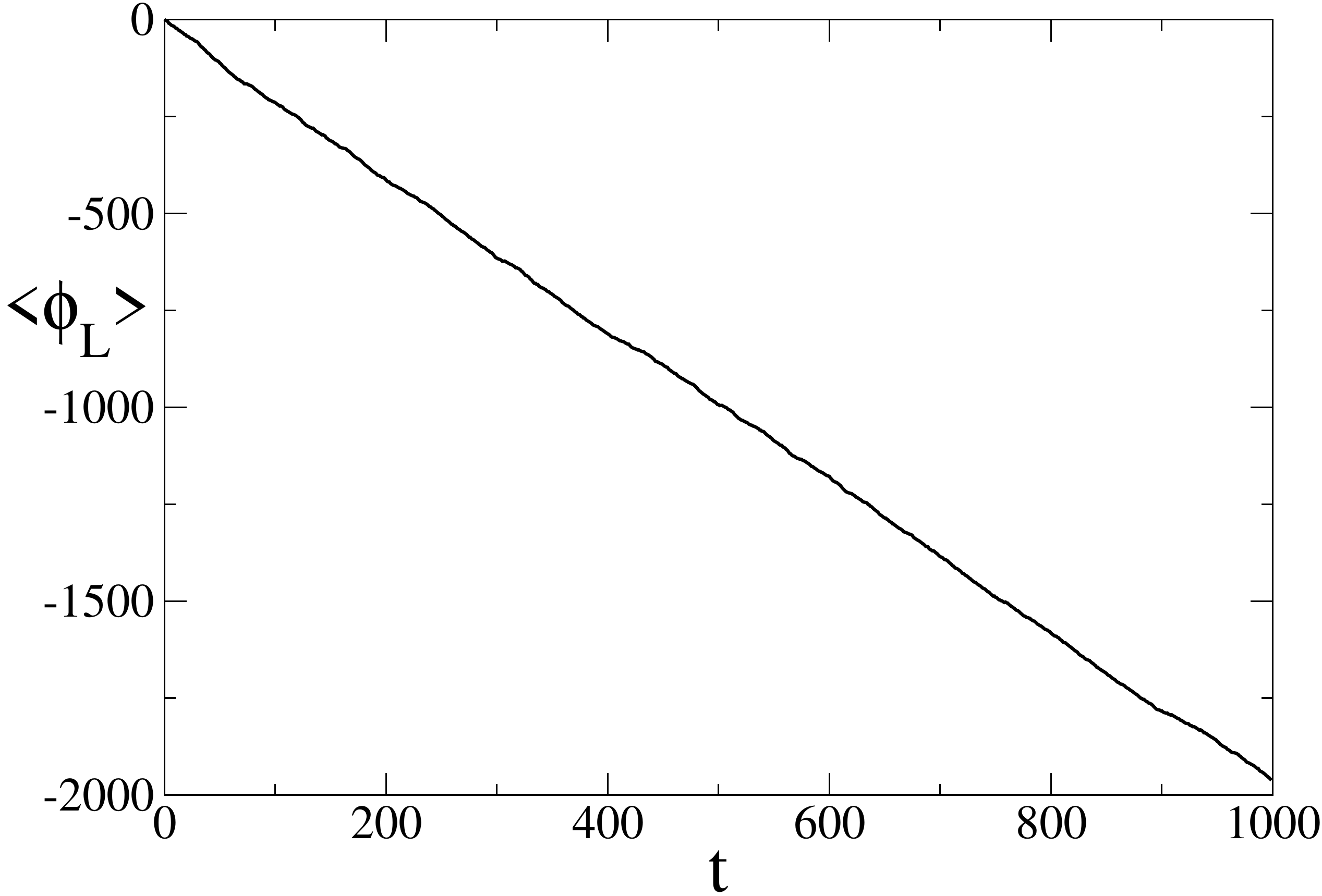}}\hfill
\caption{Plot of the time evolution of the average aggregate angle obtained from: (a) simulation of the 2D stochastic model using the Gillespie algorithm;
 (b) numerical integration of the 1D Langevin equation. For each, the parameters are $K=5000$, $b=K^2$, $q=0.5$, $p_1=20$ and $p_2=0$, 
and the average is taken over 100 runs.}
\label{Figgillandlephi}
\end{center}
\end{figure}

\subsection{Numerical integration of the Langevin equation}

The average velocity can also be calculated from direct numerical integration of the quasi-1D Langevin equation~(\ref{LEphi}). We do so using the integration
scheme~\cite{Risken}
\begin{equation}
\phi_L(t+dt)=\phi_L(t)+f(\phi_L)dt+g(\phi_L)\sqrt{dt}~\eta_\phi \;.
\end{equation}
For the same set of parameter values of previously, we find an ensemble-averaged angular displacement shown in figure~\ref{Figtyplephi}. This time we find that
$\omega_L=-1.27$, in reasonable agreement with the simulations of the full 2D diffusion. (See below for a more detailed discussion of the different methods
for estimating the current).

\subsection{Kramers' escape-rate theory}

Our final approach makes use of the transformation of the diffusion with multiplicative noise to a diffusion with additive noise described in
Section~\ref{SecQuasi1D}. This allows us to estimate the nonequilibrium current via calculations of escape rates over potential barriers as done by
Kramers~\cite{Kramers}, following closely the presentation of the method in \cite{Risken}.

The potential $\Phi(\theta)$ that we are required to calculate is
\begin{equation}
 \Phi (\theta)=-\int^{\theta}d\theta'F(\theta')\;. \label{Phitheta}
\end{equation}
From earlier discussion, one would expect $\Phi(\theta)$ to only be a true potential if $F(\theta)$ is conservative. However, using the ansatz~(\ref{Fansatz})
in~(\ref{Phitheta}) we have
\begin{equation}
 \Phi(\theta)=-\omega_\theta \theta +V(\theta)
\end{equation}
up to an irrelevant constant. We know $V(\theta)$ is periodic as it can be expressed as a Fourier series (\ref{fourierseries}).
Therefore in the domain of $\theta$, $V$ will be monotonically shifted  by $\omega \theta$, which we interpret as our potential $\Phi(\theta)$.

Given the form of $f$ and $g$ derived for this model it is not possible to perform the transformation from $\phi$ to $\theta$ in (\ref{phitothetadef}),
or compute the integral in (\ref{Phitheta}) to obtain $\Phi$. By approximating the function $g$ with a mathematically tractable function we are
able to make some progress in analytically approximating these required expressions. The technical details of this can be found in~\ref{AppPhi}.
\begin{figure}[t]
\begin{center}
 \subfigure[\hspace{1mm}$p_1=20$, $p_2=0$]{\label{FigPhip1}\includegraphics[width=0.22\linewidth]{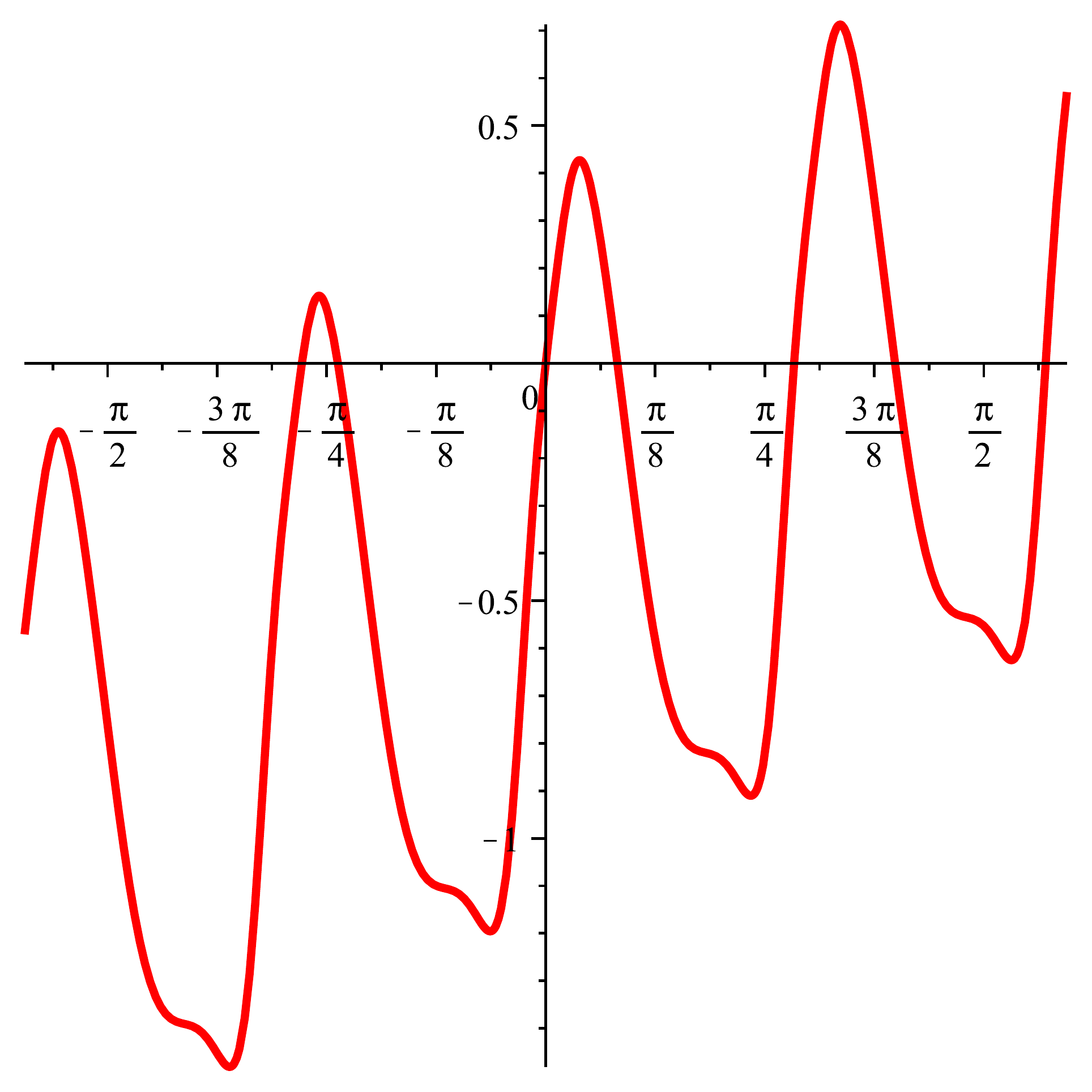}}\hspace{10mm}
\subfigure[\hspace{1mm}$p_1=0$, $p_2=0$]{\label{FigPhieqm}\includegraphics[width=0.22\linewidth]{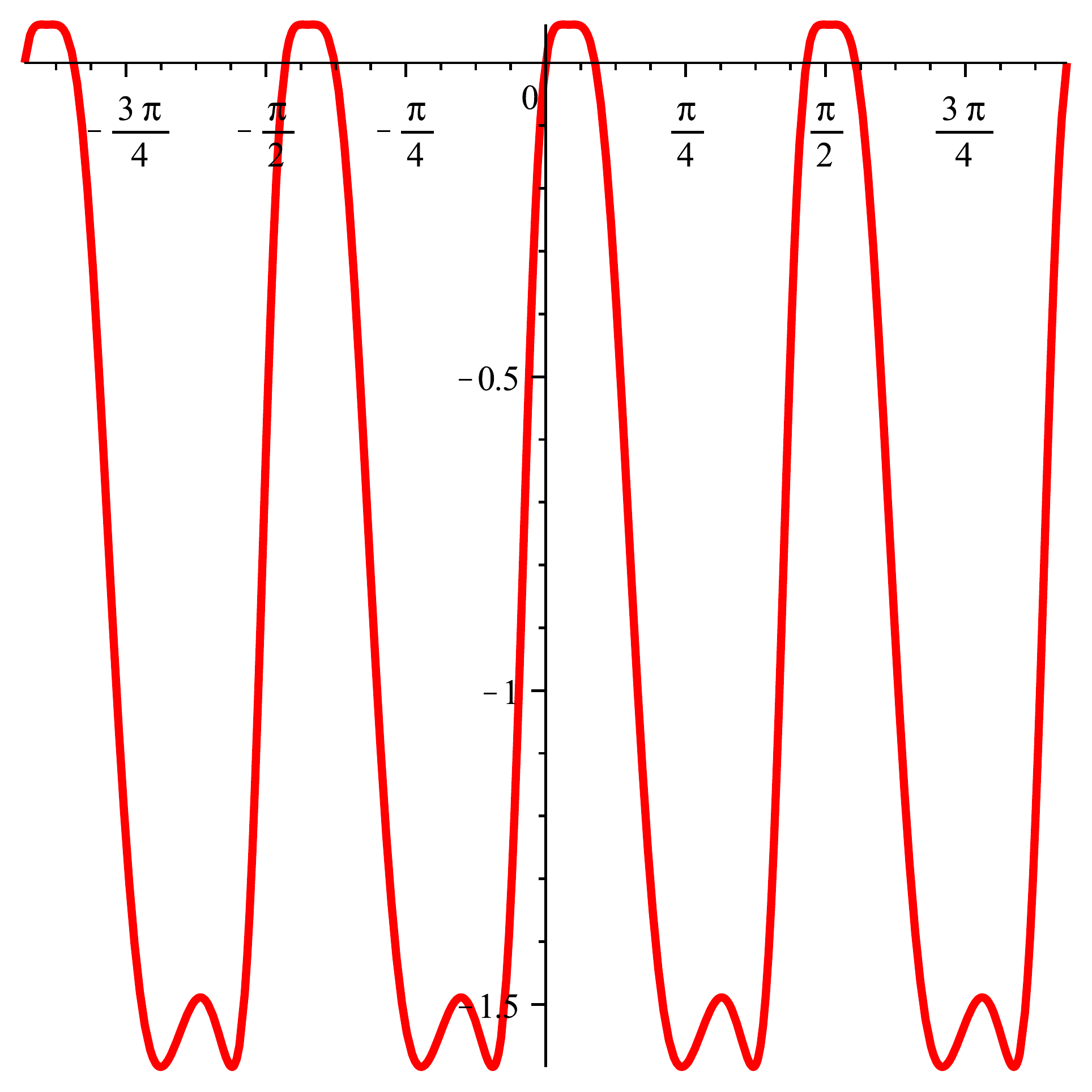}}\hspace{10mm}
  \subfigure[\hspace{1mm}$p_2=0$, $p_2=20$]{\label{FigPhip2}\includegraphics[width=0.22\linewidth]{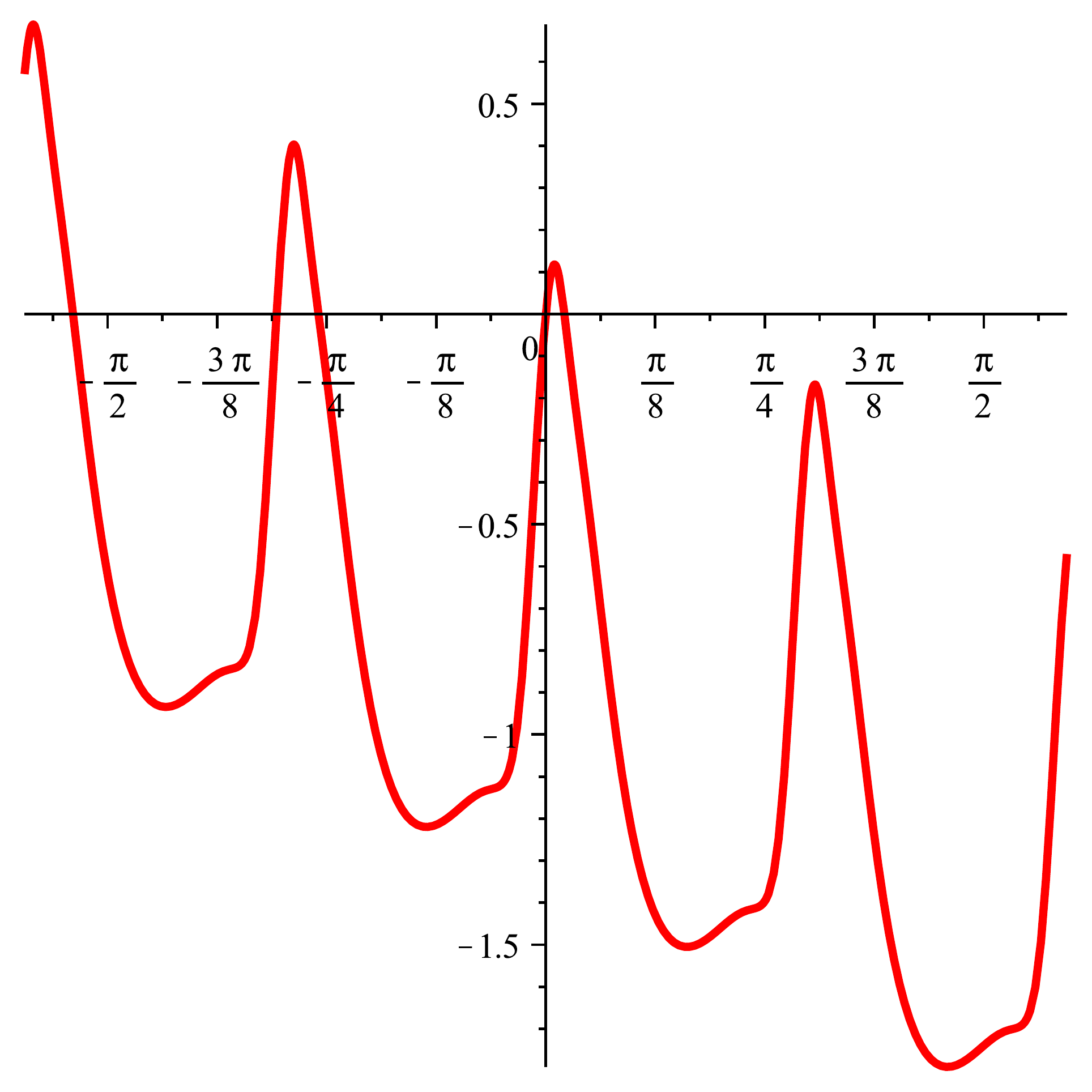}}\hfill
\caption{$\Phi(\theta)$ for different values of the nonequilibrium parameters $p_1$, $p_2$ with $q=0.5$.}
\label{FigPhi}
\end{center}
\end{figure}

In figure~\ref{FigPhi} are typical examples of $\Phi(\theta)$ according to different values of the non-equilibrium parameters $p_1$ and $p_2$.
The thermal equilibrium condition $p_1=p_2$ is shown in figure~\ref{FigPhieqm}. In this case there is no current flowing and the potential $\Phi(\theta)$ 
is periodic. However in the nonequilibrium cases shown in figures~\ref{FigPhip1} and \ref{FigPhip2} we see that $\Phi(\theta)$ is not periodic.
Though as the domain of of the potential is only $\theta \in[-\theta_0,\theta_0]$, as defined in~\ref{AppPhi}, it is only important that the upper and 
lower limits of the domain are the same to satisfy the periodic boundary conditions of the system. The absolute value of a potential is not important
for deriving the force for it, so we do not require $\Phi(\theta)$ to be periodic to regard it as being a potential.

Each of the potentials in figure~\ref{FigPhi} are characterised by having a minimum between two barriers. We can derive the escape rate of a 
particle over these barriers; here we will do so for the right side barrier of the potential displayed in figure~\ref{FigPhip2}.
We denote the position of the minimum in the well by $\theta_{W}$, the top of the barrier by $\theta_{B}$
and the next minimum to the right of the barrier by $A$. For the FPE~(\ref{FPtheta}) in a nonequilibrium steady state the probability current
\begin{equation}
 J=-\partial_\theta \Phi_\theta(\theta)Q_S(\theta)-D\partial_\theta Q_S(\theta)
\end{equation}
can be written as
\begin{equation}
 J=-De^{-\Phi_\theta/D}\frac{\partial}{\partial\theta}\bigg[e^{\Phi_\theta/D}Q_S\bigg]\;. \label{SKramers0}
\end{equation}
Integrating between $\theta_{W}$ and $A$ we have
\begin{equation}
 J\int_{\theta_W}^{A} \! {\rm d} \theta\ e^{\Phi_\theta/D}=D\left[e^{\Phi_\theta(\theta_{W})/D}Q_S(\theta_{W})-e^{\Phi_\theta(A)/D}Q_S(A)\right]\;. \label{SKramers1}
\end{equation}
If the barrier height $\Delta \Phi=\Phi(\theta_{B})-\Phi(\theta_{W})$ is much greater than the diffusion $D$ then the particle is far more
likely to be found in the well about $\theta_{W}$. This means we neglect the second term in the square brackets in~(\ref{SKramers1}), giving
\begin{equation}
J = \frac{D\ Q_S(\theta_W)\ {\rm e}^{\Phi(\theta_W)/D}}{ \int_{\theta_W}^{A} {\rm d} \theta\ e^{\Phi_\theta/D}} \;. \label{SKramers2}
\end{equation}
The current can be expressed as the probability $p$ of being in the well at $\theta_{W}$ multiplied by the escape rate $r$ from the well. 
Taking $\theta_1<\theta_{W}<\theta_2$ to define the domain of the well we write
\begin{equation}
 p=\int_{\theta_1}^{\theta_2}\! {\rm d} \theta\ Q_S(\theta)\;. \label{pKramers}
\end{equation}
From~(\ref{SKramers0}) the stationary distribution is
\begin{equation}
 Q_S(\theta)=Ne^{-\Phi(\theta)/D}-Je^{-\Phi(\theta)/D}\int^{\theta}\dint{\theta'}\frac{e^{\Phi(\theta')/D}}{D}\;.
\end{equation}
As we are interested in the stationary distribution in the well we introduce $Q_S(\theta_{W})$ by eliminating the normalisation constant $N$:
\begin{equation}
\fl  Q_S(\theta)=Q_S(\theta_{W})e^{-[\Phi(\theta)-\Phi(\theta_{W})]/D}-\frac{Je^{-\Phi(\theta)/D}}{D}\left[\int^{\theta}\dint{\theta'}e^{\Phi(\theta')/D}
-\int^{\theta_W}\dint{\theta'}e^{\Phi(\theta')/D}\right] \label{Qsmin}
\end{equation}
Using this we can now write 
\begin{equation}
 p=Q_S(\theta_W)e^{\Phi(\theta_W)/D}\int_{\theta_1}^{\theta_2}\dint{\theta}e^{-\Phi(\theta_W)/D}
\end{equation}
where the contribution from the square bracket terms in (\ref{Qsmin}) is negligible when considering the contribution from the domain of the well.
Combining this with~(\ref{SKramers2}) we can express the escape rate as 
\begin{equation}
 \frac{1}{r}=\frac{p}{J}=\frac{1}{D}\int_{\theta_1}^{\theta_2} \! {\rm d}\theta\ e^{-\Phi(\theta)/D}\int_{\theta_{W}}^{A} \! {\rm d} \theta\ e^{\Phi(\theta)/D}\;.
\label{rKramers}
\end{equation}

We Taylor expand each integrand in the above expression, the first about $\theta_{W}$, the second about $\theta_{B}$:
\begin{eqnarray}
\Phi(\theta)&\approx \Phi (\theta_{W})+ \frac{\Phi''(\theta_{W})}{2}(\theta-\theta_{W})^{2} \\
\Phi(\theta)&\approx \Phi (\theta_{B})- \frac{|\Phi''(\theta_{B})|}{2}(\theta-\theta_{B})^{2}\;.
\end{eqnarray}
Using these second order expansions, we can extend the boundaries of each integral to $\pm \infty$, meaning that to both integrals in~(\ref{rKramers})
are now Gaussian. Computing them leads to the final expression for the escape rate
\begin{equation}
 r\approx \frac{1}{2\pi}\sqrt{\Phi''(\theta_{W})|\Phi''(\theta_{B})|}~e^{-\Delta\Phi /D}\;. \label{r}
\end{equation}
To find the value of these escape rates, we numerically compute the value of the second derivatives at the minima and maxima after explicitly constructing
the potential $\Phi(\theta)$ using the mathematical software Maple (TM).

Denoting by $r^-$ and $r^+$ the escape rate for the left and right barrier respectively, we express the  average change in $\theta$ due to hopping over
the right or left barrier in a time $\tau$ as
\begin{equation}
 \langle \delta \theta \rangle=(r^+\Delta_\theta-r^-\Delta_\theta)\tau
\end{equation}
where $\Delta_\theta=2\theta_0$ is the distance between wells, i.e. the period of the system. In the limit $\tau \rightarrow 0$ this gives
the average angular velocity
\begin{equation}
 \omega_K=\Delta_\theta(r^+-r^-)\;.
\end{equation}
It is clear from (\ref{r}) that a current exists due to the difference in the height of the two barriers. When the system is in thermal equilibrium as in
figure~\ref{FigPhieqm} there is no current as we are equally likely to hop left or right around one circuit.
With the parameter set used before, and setting $D=1$, we find $\omega_K=-0.83$. Again, this value is in reasonable agreement with those previously obtained by
other means.

\subsection{Comparison of the three methods}

To better understand how well the currents obtained via these different approaches correspond with each other, we compare in Figure~{\ref{Figomegadiffp}
these measures for different values of the $p$ parameter which controls the extent to which detailed balance is violated. We see that all three
measurements obey the same qualitative trend, and remain within an order of magnitude of each other.
\begin{figure}[t]
\begin{center}
\includegraphics[width=0.45\linewidth]{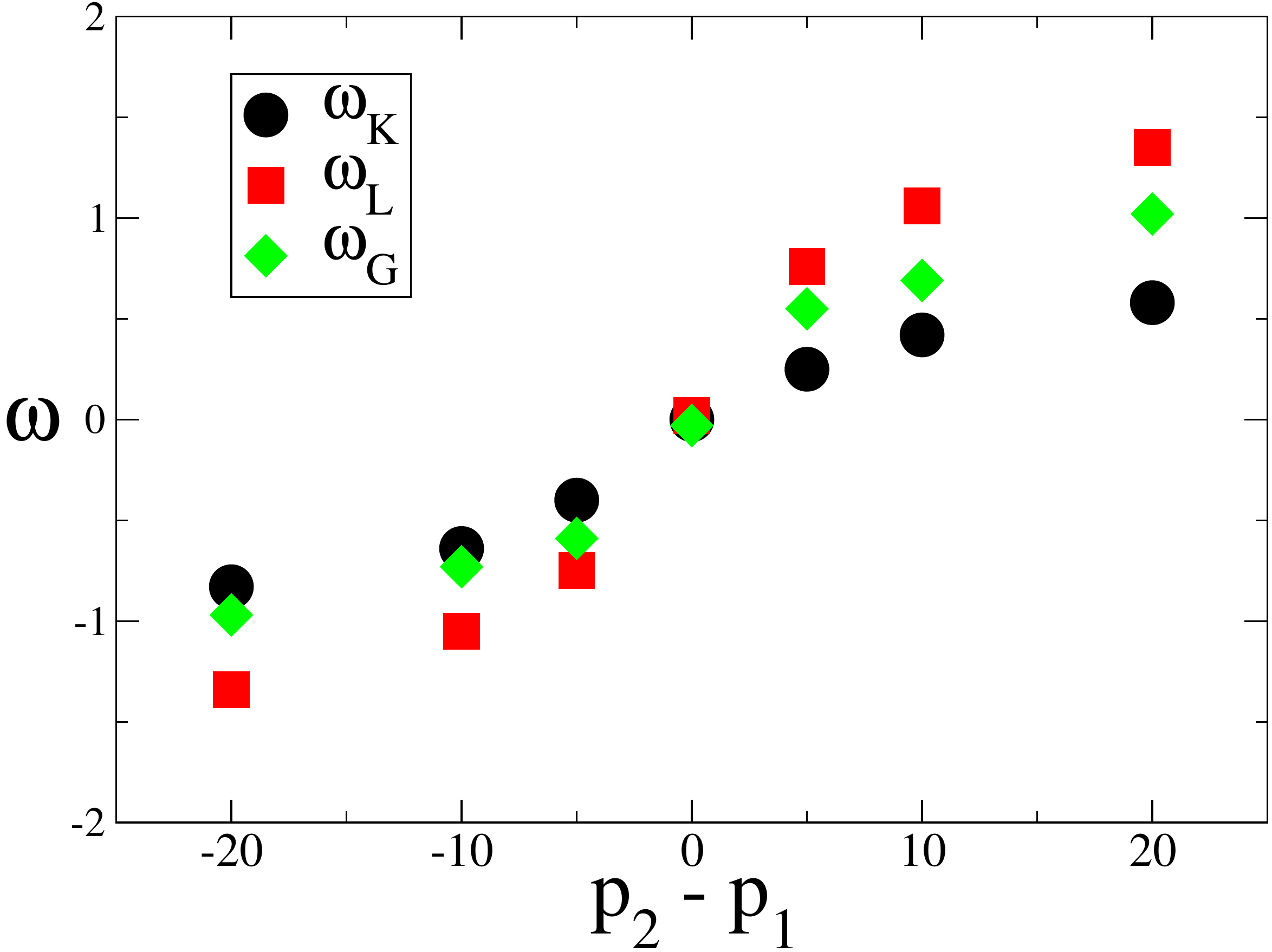}
\caption{Overlay of the different measurements of $\omega$ using the three distinct methods of section \ref{SecMeasure} plotted against a relative
measure of the nonequilibrium current $p_2-p_1$. For each data point either $p_1$ or $p_2$ is zero. }
\label{Figomegadiffp}
\end{center}
\end{figure}

This validates our principal approximation to reduce the full two dimensional system to one dimension, that of the polar angle $\phi$,  by neglecting radial
diffusion. In particular, we note that the 1D criterion for a non-equilibrium steady state $p_1=p_2$ is borne out by simulations of the full 2D dynamics. The
difference in the dynamic measurements, $\omega_L$ is larger than $\omega_G$, is understandable as while we neglect any radial diffusion in the 1D treatment, it
is still present in the simulations, as witnessed in figure~\ref{Figtypstate}. We expect time spent diffusing radially to slow the rate of polar diffusion. 

The main assumption of the Kramers escape rate calculation is that the ratio of the barrier size $\Delta\Phi$ to the diffusion constant $D$ is very large.
In practice, we set $D=1$, and it is not possible to tune the model to allow us to independently control the barrier height and the diffusion strength.
Typically, we find that the ratio is  $2<\Delta\Phi/D<3$, which is a likely cause of the quantitative discrepancy between this method 
of obtaining $\omega$ with the other two.

\section{Discussion and Conclusions}
\label{SecDisandCon}

In this work, it has been our aim to understand the conditions under which a nonequilibrium steady-state current in a multi-species population dynamics model
may be visible macroscopically in the form of population abundance cycles. The simplest candidate for such a system has an attractive fixed point and an absence
of detailed balance. Here we found that cyclicity is not evident on the timescale of a single (probability) cycle in a single realisation of the
dynamics. One must either perform an average over long times or an large ensemble to uncover this weak systematic cyclicity.

One situation that allows these currents to be observed in single realisations of the dynamics has a deterministic dynamics that is neutral on some
closed manifold in configuration space. By this we mean that all deterministic forces vanish on the manifold, and that the manifold itself is attractive.
A nonequilibrium steady state can then be manifested as a systematic orbit around the manifold.
 Although, the specific interactions that yield such a structure 
are somewhat contrived from a purely biological viewpoint, this work highlights some interesting features of stochastic dynamical systems 
from the perspective of nonequilibrium statistical mechanics.

\begin{figure}[t]
\begin{center}
\includegraphics[width=0.45\linewidth]{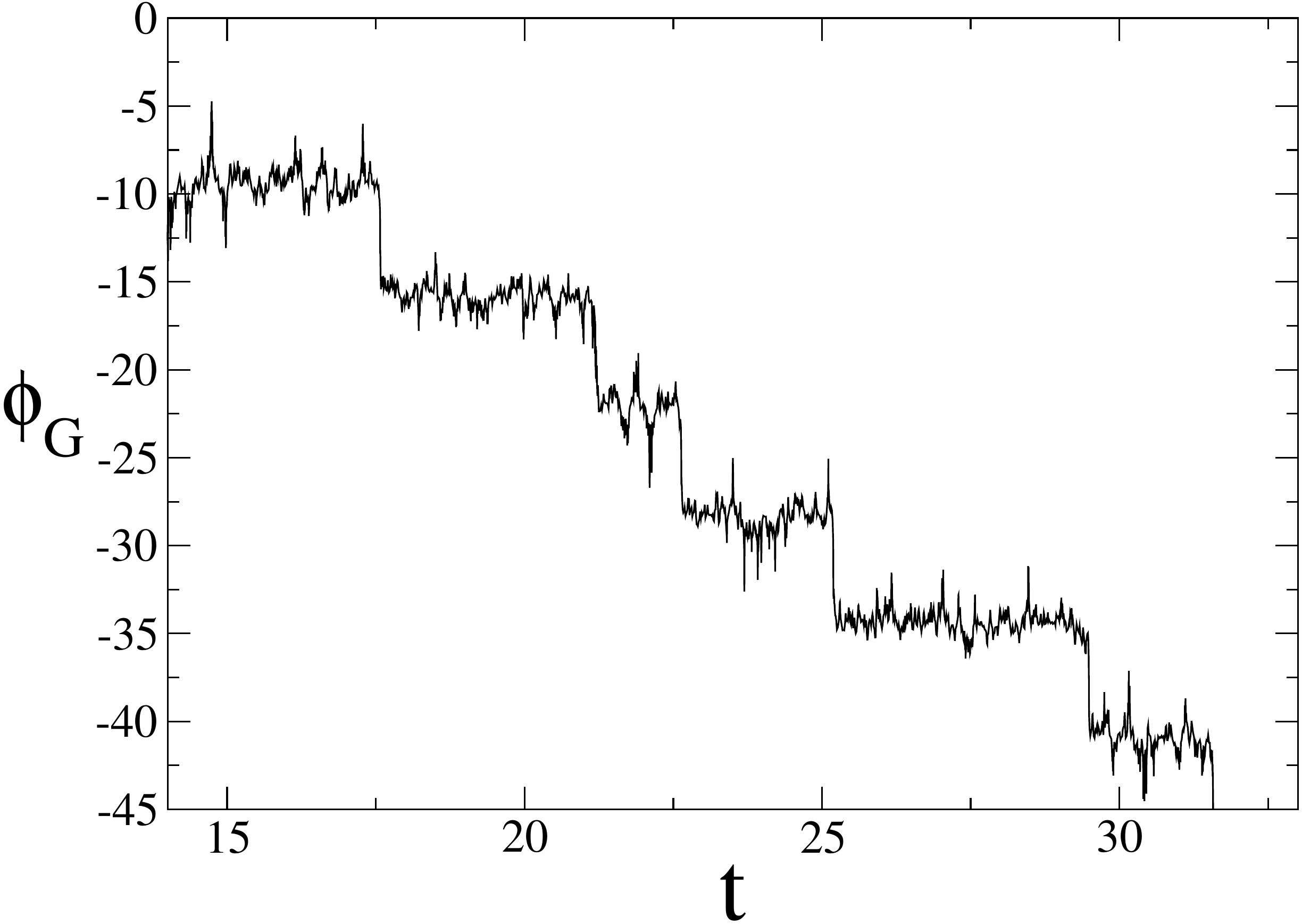}
\caption{A single realisation of $\phi_G$ obtained by simulation of the stochastic dynamics with parameters:
$K=5000$, $b=K^2$, $q=0.5$, $p_1=20$, $p_2=0$.}
\label{Figgillsteps}
\end{center}
\end{figure}

Most notably, the stochastic trajectories along the manifold are somewhat complex, as figure~\ref{Figgillsteps} shows.
 We see that the system tends to  diffuses over a small region of the phase space before sharply jumping a distance of about ${}\approx2\pi$ radians, 
i.e.~a full circuit of the manifold. The origin of this motion can be understood from the potential picture after mapping to additive (thermal) noise as described
 in section~\ref{SecQuasi1D}.  After transforming the multiplicative noise, the deterministic equations acquire additional terms that can be interpreted
 as a constant driving force acting in a periodic potential with multiple maxima and minima on the manifold.  The dynamics will reside for some time
 in a potential minimum before escaping over one of the barriers to a neighbouring minimum.  Since the potential is periodic, one eventually returns 
to the same minimum (hence the $2\pi$ jump). The driving force leads to an asymmetry in the barrier heights, which in turn yields a systematic current
 in one direction around the circle (i.e., a `cycle', albeit not a strictly periodic one).

This analysis leads us to believe that in general the closed manifold does not have 
to be neutrally stable in order to observe this cyclical behaviour, as long as the manifold itself is attractive from the outside. In such an instance, the non-zero angular velocity generated by the probability current 
would still drive the system over potential barriers, but now they barriers are due to the deterministic forces acting on the manifold. 
While the time scale for passing over a barrier would increase markedly, quasicycles would still be generically observable in a system with athermal noise
whose deterministic dynamics evolve to a closed manifold.

The observation of quasicycles due to a noise-induced non-conservative driving force highlights the importance of the form of the noise that is manifest in these stochastic population models. 
In the $\phi$ Langevin equation~(\ref{LEphi}) the non-equilibrium processes given in (\ref{newprocesses}) only appear in the diffusion term $g$. 
Therefore if one naively assumed the noise to be thermal, i.e. $g$ is constant, integrating up the drift force $f$ would yield a periodic potential 
similar to the one displayed in figure~\ref{FigPhieqm}, and one would conclude no current is flowing. We see that stochastic effects alone are responsible 
for a current to flow in the system, keeping the system out of equilibrium.

In the context of modelling population dynamical systems more generally, our findings further showcase the advantages of taking an IBM approach to modelling 
finite population dynamics. The deterministic and stochastic contributions to the dynamics can be derived from the defined stochastic processes of the model,
 so one can analytically understand and also control the effects of noise. We have found that finite size populations in which the dynamics are non-neutral,
 never relax fully to equilibrium, but instead inhabit a steady state where their is a thermal bias in the fluctuations due to the presence of a non-equilibrium
 probability current.

In this work we resorted to a number of approximations to compute the steady state current in the dynamical system.  The first of these was a reduction to a 
single-coordinate description by disregarding one of the degrees of freedom in the system.  Sophisticated methods have been applied to integrate out this degree
 of freedom in the context of quasi-neutral diffusion along an open interval \cite{ParsonsQuince} (as opposed to one that is closed/periodic, as here). 
It would be of interest to see if similar methods can be applied to determine what is lost in such a dimensional reduction, as this may be of utility 
in understanding high-dimensional stochastic dynamical systems more generally.  Moreover, we made various approximations in order to apply 
Kramers' escape rate theory to diffusion on the manifold: it may be that more direct approaches to estimating the current in such systems can be found. 
 Finally, and more generally, it would be interesting to establish if there are other ways that a nonequilibrium current may enter into the macroscopic dynamics
 of a stochastic population dynamical system in ways that are not immediately evident from their deterministic counterparts.

\section*{Acknowledgements}
This work has made use of the resources provided by the Edinburgh Compute and Data Facility (ECDF). (http://www.ecdf.ed.ac.uk/). 
The ECDF is partially supported by the eDIKT initiative (http://www.edikt.org.uk). We thank the EPRSC (D.I.R.) and RCUK (R.A.B.) for financial support.

\appendix
\section{The polar Fokker-Planck equation}
\label{AppFP}
\setcounter{section}{1}
To derive a Fokker-Planck equation for the evolution of the probability density $P(\phi,t)$ we terminate the Kramers Moyal forward expansion \cite{Risken}
\begin{equation}
 \frac{\partial P(\phi,t)}{\partial
t}=\sum_{n=1}^{\infty}\frac{(-1)^{n}}{n!}\left(\frac{\partial}{\partial\phi}\right)^n\alpha_n(\phi)P(\phi,t) \label{KMexp} \\
\end{equation}
at the second term. This truncation is valid if $\alpha_1$ and $\alpha_2$ are the only non-zero jump moments as defined by
\begin{equation}
 \alpha_n=\lim_{\tau \rightarrow 0}\frac{\langle[\phi(t+\tau)-\phi(t)]^n\rangle}{\tau}\equiv \lim_{\tau \rightarrow 0} 
\frac{\langle (\delta \phi)^n\rangle}{\tau}\;. \label{jumpmomdef}
\end{equation}
 The moments of $\delta \phi$ can be obtained from the moments of $\delta x_A$ and $\delta x_B$ by the relation
\begin{equation}
 \delta \phi(\delta x'_{A}, \delta x'_{B}|x'_{A},x'_{B})=\tan^{-1}\left(\frac{x'_{B}+\delta x'_{B}}{x'_{A}
+\delta x'_{A}}\right)-\tan^{-1}\left(\frac{x'_{B}}{x'_{A}}\right)\;. \label{deltaphi}
\end{equation}
Taylor expanding to second order in $\delta x'_{A}$ and $\delta x'_{B}$ about $\delta x'_{A}=\delta x'_{B}=0$ and keeping terms up to order $\delta^2$ yields
\begin{eqnarray}
  \delta \phi=&-\frac{x'_{B}}{r^{2}}(\delta x'_{A})+\frac{x'_{A}}{r^{2}}(\delta x'_{B}) \nonumber \\
&+\frac{x'_{A}x'_{B}}{r^{4}}\left(\delta (x'_{A})^{2}-(\delta x'_{B})^{2}\right)+\frac{(x'_{B})^{2}-(x'_{A})^{2}}{r^{4}}(\delta x'_{A}\delta x'_{B}) \;. \label{Taylorphi}
\end{eqnarray}
Applying the polar transformation (\ref{polarcoord}) and averaging yields
\begin{eqnarray}
\hspace{5mm}\langle \delta \phi\rangle&=\frac{b}{K^2}\left[\frac{\sin(2\phi)}{2r_0^2}\bigg(M_{2,0}-M_{0,2}\bigg)\right]\tau  \label{1stmom}\\
\nonumber \\
\langle (\delta \phi)^2 \rangle&=\frac{b}{K^2}\left[\frac{\sin^2(\phi)M_{2,0}+\cos^2(\phi) M_{0,2}}{r_0^2}\right]\tau \label{2ndmom} \;.
\end{eqnarray}
where we use the fact that deterministically we are at a fixed point so $\langle \delta x_A\rangle=\langle \delta x_B \rangle=0$.

For the higher order jump moments we can infer from~(\ref{Mcartmoms}) and~(\ref{M12cartmoms}) that to leading order in $K$
\begin{equation}
 \langle(\delta \phi)^n\rangle\sim bK^{-n}G_n(x_A,x_B)\tau, \hspace{5mm}n\geq3
\end{equation}
where $G_n$ are functions of the intensive variables, independent of $K$. Using~(\ref{jumpmomdef}), we have from the above expressions for the moments that
\begin{eqnarray}
 \alpha_1,\alpha_2 &\sim \lim_{b \rightarrow \infty}\frac{b}{K^2}  \\
\hspace{6mm}\alpha_n &\sim \lim_{b \rightarrow\infty}\frac{b}{K^n},\hspace{5mm}n\geq3\;.
\end{eqnarray}
The change in limit is an observation that the rates defined in (\ref{model}) scale with $b$ and so the time $\tau$ until an event happens scales like $1/b$.
This means that in order to have a non-zero first and second jump moment with higher order jump moments vanishing we should choose $b=K^2$.

\section{Approximating $\theta(\phi)$ and $\Phi(\theta)$}
\label{AppPhi}
\setcounter{section}{2}

We are faced with the task of computing 
\begin{eqnarray}
 \Phi(\theta)&=-\int^{\theta}\dint{\theta'}F(\theta')\\
&=-\sqrt{2D}\int^{\theta}\dint{\theta'}\left(\frac{f(\phi)}{g(\phi)}-\frac{1}{2}\partial_{\phi}g(\phi)\right)\bigg|_{\phi=\phi(\theta')}\;.
\end{eqnarray}
Applying the same change of variable as in (\ref{a0phitotheta1}), (\ref{a0phitotheta2}) we have
\begin{equation}
\Phi(\theta)=
-\sqrt{2D}\int^{\phi(\theta)}\dint{\phi}\left(\frac{f(\phi)}{g^{2}(\phi)}-\frac{1}{2}\frac{\partial_{\phi}g(\phi)}{g(\phi)}\right)\;. \label{Phiintg}
\end{equation}
The form of $g(\phi)$ as it is given in~(\ref{g}) makes the above integral intractable. We therefore make the approximation
$G(\phi)\approx G_{g}(\phi)= 1/g(\phi)$ where $G$ has the form
\begin{equation}
 G(\phi)=c_{G}+A_{G}\cos(\phi-b_{G}) \;. \label{Gapprox}
\end{equation}
The phase $b_{G}$ is set to by matching numerically the position of the first extremum of $G_{g}$ and $G$ for $\phi>0$. The other two parameters are set by
making the maximum and minimum values of $G_{g}$ and $G$ the same. We illustrate this approximation for a typical $g$ in figure~\ref{gandG}.
\begin{figure}[t]
\begin{center}
\includegraphics[width=0.4\linewidth]{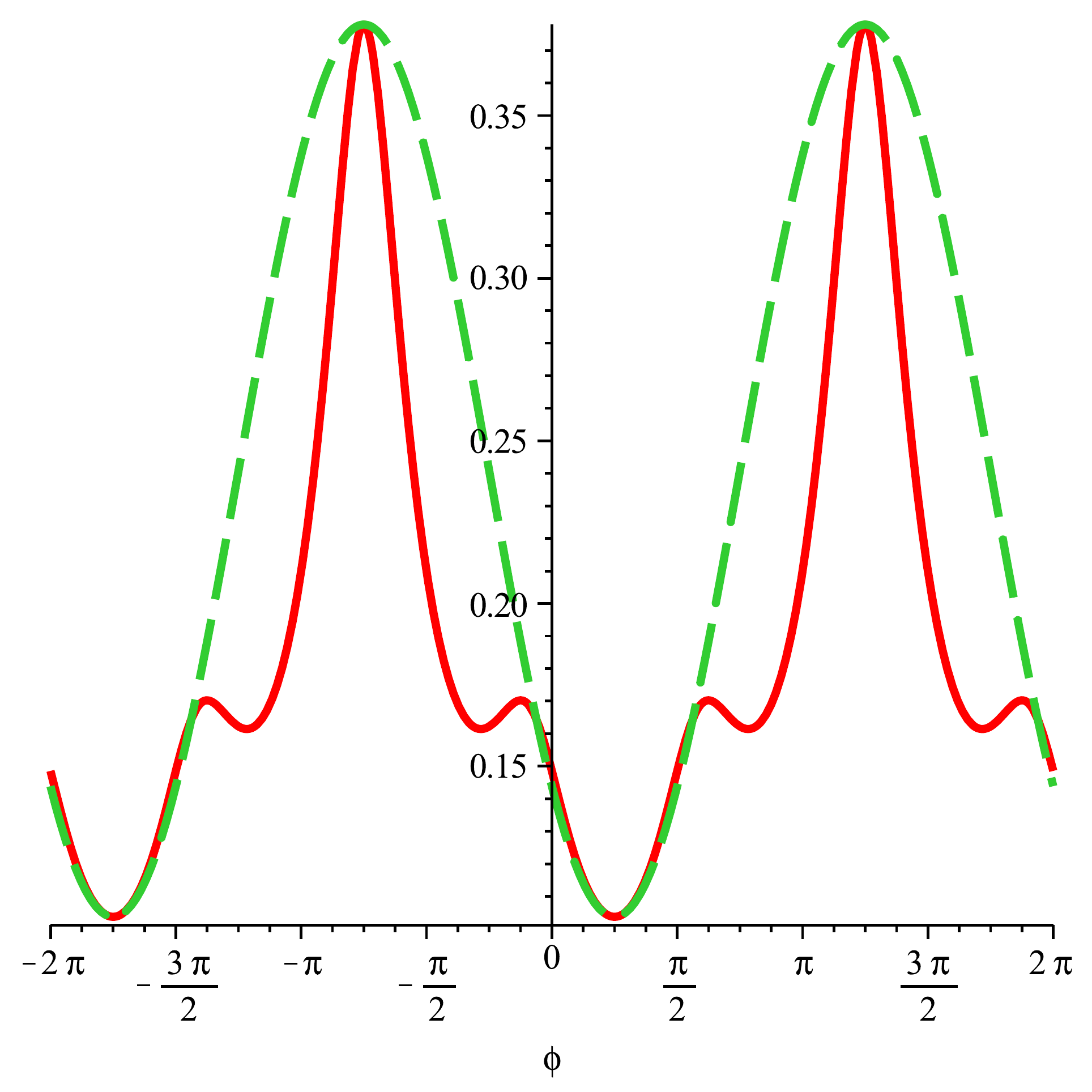}
\caption{The inverted diffusion term $1/g$ (red, solid) and its approximation $G$ (green, dashed).}
\label{gandG}
\end{center}
\end{figure}
 Applying the approximation $G(\phi)\approx 1/g(\phi)$ our expression~(\ref{Phiintg}) becomes
\begin{equation}
\Phi_\theta(\theta)=-\sqrt{2D}\int^{\phi(\theta)}\dint{\phi} \left(f(\phi)G(\phi)^2+\frac{1}{2}\frac{\partial_{\phi}G(\phi)}{G(\phi)}\right)\;. \label{finalPhiint}
\end{equation}

It is possible to numerically integrate this integral by computer. However, to evaluate it in the $\theta$ coordinate we must have an analytical form of
$\phi(\theta)$ to substitute into it.

The transformation from $\theta$ to $\phi$ in~(\ref{phitothetadef}) does not have a closed form for the exact $g(\phi)$ derived for this model
so we approximate it by
\begin{equation}
 \theta(\phi)\approx\theta_G(\phi)= \sqrt{2D}\int_{0}^{\phi}d\phi'G(\phi')
\end{equation}
which using~(\ref{Gapprox}) gives
\begin{equation}
 \theta_G(\phi)=\sqrt{2D}(c_{G}\phi -A_{G}\sin(\phi-b_{G})-A_{G}\sin(b_{G}))\;.
\label{thetaG}
\end{equation}
This functional relation is shown in figure~B\ref{ThetaGfig}. What we really require is the inverted form of this relation, giving us $\phi$
in terms of $\theta$. The inversion is not possible, however we can approximate it via 
\begin{equation}
 \theta_{t}=\sqrt{2D}\int_{0}^{\phi}\frac{\rm d\phi}{c_{t}+A_{t}\cos(\phi)}\;.
\end{equation}
This integral has the solution
\begin{equation}
 \theta_{t}(\phi)=\frac{2\sqrt{2D}}{c_{t}\sqrt{1-a_{t}}}\tan^{-1}\left(\frac{1-a_{t}}{\sqrt{1-a_{t}^{2}}}\tan\left(\frac{\phi}{2}\right)\right)\label{thetaGtan}
\end{equation}
where $c_{t}=A_{t}/c_{t}$. We need to approximate the values $c_{t}$ and $a_{t}$ to give a good fit with the form of $\theta(\phi)$ from~(\ref{thetaG}).

We do this by matching the linear gradient and the curvature near $\phi=0$ of $\theta_G(\phi)$ and $\theta_{t}(\phi)$.
The slope we match by finding the straight line gradient between two points that lie on $\theta_{t}(\phi)$ and making it equal to slope of the constant term 
in~(\ref{thetaG}). Choosing the two points $\phi=-\pi$ and $\phi=\pi$ we have from~(\ref{thetaGtan}) that
\begin{equation}
 \theta_{t}(\pm \pi)=\pm\frac{\sqrt{2D}~\pi}{c_{t}\sqrt{1-a_{t}^{2}}}\;. 
\end{equation}
Matching the gradients:
\begin{equation}
 \frac{\theta_{t}(\pi)-\theta_{t}(-\pi)}{2\pi}=\sqrt{2D}~c_{G}
\end{equation}
yields
\begin{equation}
 c_{G}=\frac{1}{c_{t}\sqrt{1-a_{t}^{2}}}\;. \label{coeffmatch1}
\end{equation}

To match the curvature near the origin we rearrange~(\ref{thetaGtan}) to
\begin{equation}
 \tan\left(\frac{c_{t}\sqrt{1-a_{t}^{2}}}{\sqrt{2D}}~\frac{\theta_{t}}{2}\right)=\frac{1-a_{t}}{\sqrt{1-a_{t}^{2}}}\tan \left(\frac{\phi}{2}\right) \;.
\end{equation}
 Substituting $\theta_{G}$ for $\theta_t$ using~(\ref{thetaG}) gives
\begin{eqnarray}
 \tan\left(\frac{c_{t}\sqrt{1-a_{t}^{2}}}{\sqrt{2D}}~\frac{\theta_G}{2}\right)= \nonumber\\
\tan\left(\frac{c_{t}\sqrt{1-a_{t}^{2}}}{2\sqrt{2D}}~\bigg[\sqrt{2D}(c_{G}\phi-A_{G}\sin(\phi-b_{G})-A_{G}\sin(b_{G}))\bigg]\right) \;.
\end{eqnarray}
\begin{figure}[t]
\begin{center}
 \subfigure[\hspace{1mm}$\theta_G(\phi)$]{\label{ThetaGfig}\includegraphics[width=0.32\linewidth]{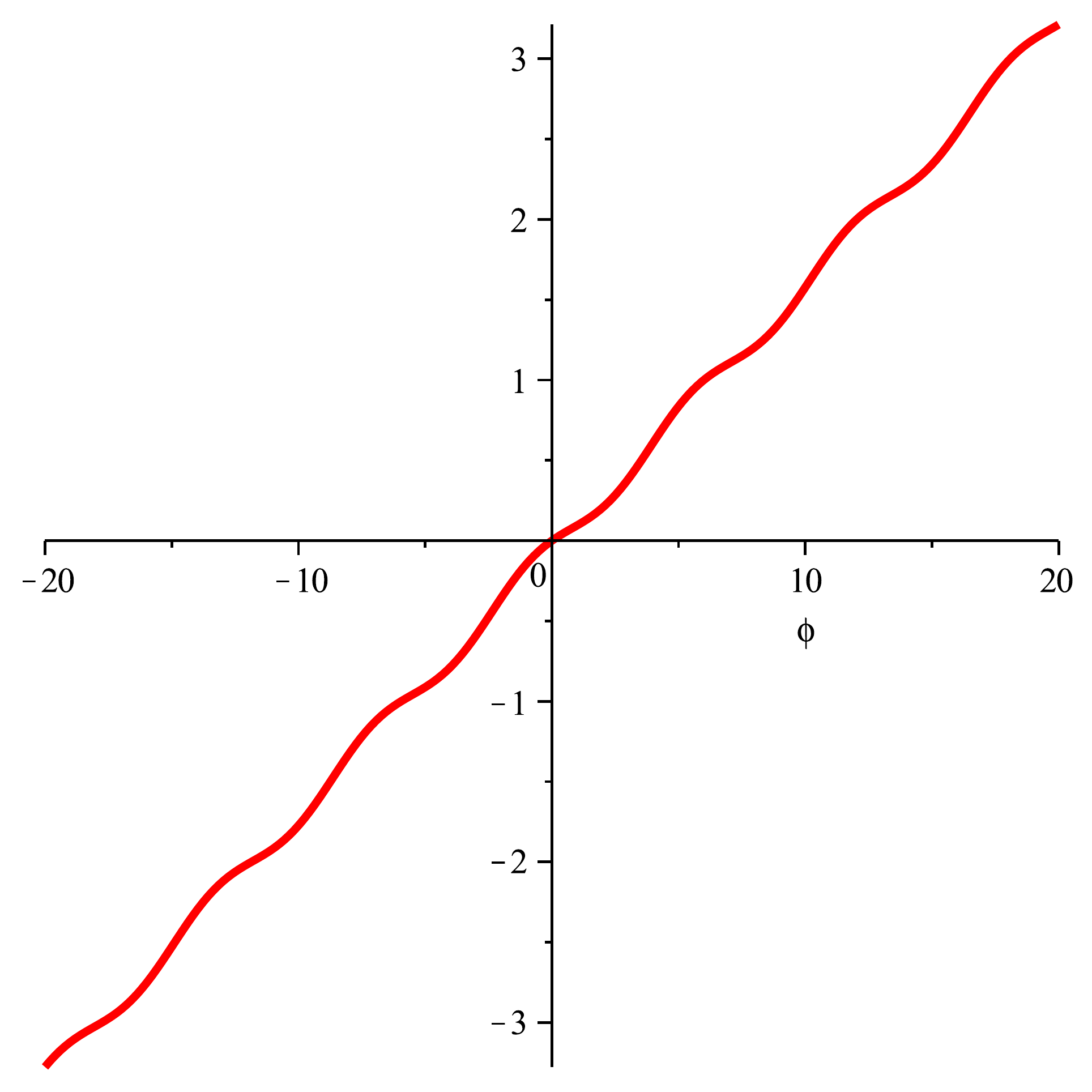}}\hspace{10mm}
  \subfigure[\hspace{1mm}$\theta_G(\phi)$ and $\theta_{t}(\phi)$]{\label{ThetaGandGappfig}\includegraphics[width=0.32\linewidth]{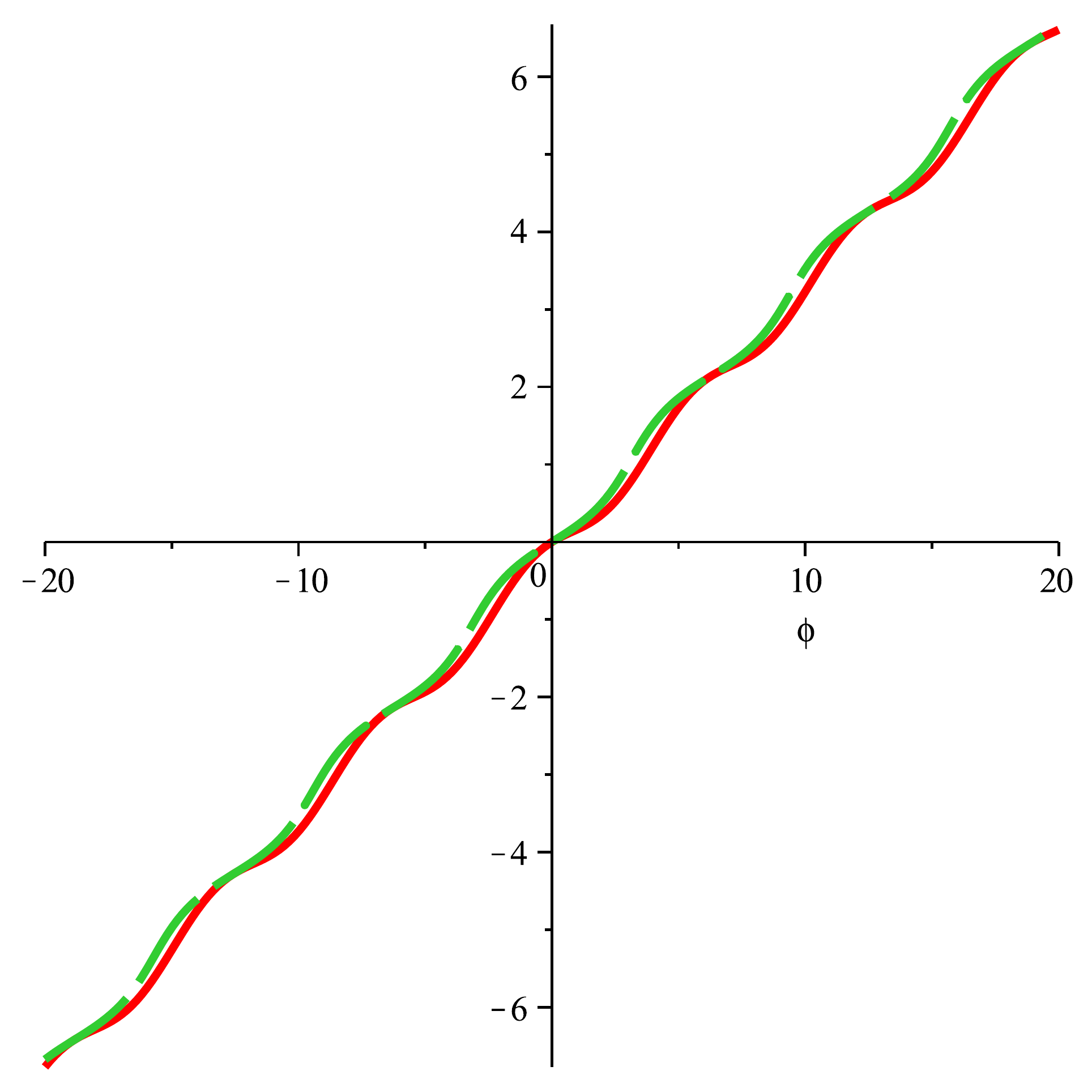}}\hfill
\caption{Comparison of the two approximations of the transformation $\theta_G(\phi)$ (red, solid) and $\theta_{t}(\phi)$ (green, dashed).}
\label{newpots}
\end{center}
\end{figure}
Now setting the above two expressions equal and Taylor expanding about $\phi=0$ we have
\begin{equation}
 \frac{1-a_{t}}{\sqrt{1-a_{t}^{2}}}~\frac{\phi}{2}=\frac{c_{t}\sqrt{1-a_{t}^{2}}}{2}\bigg(c_{G}-A_{G}\cos(b_{G})\bigg)\phi
\end{equation}
which  upon matching coefficients gives
\begin{equation}
 \frac{1}{c_{t}(1+a_{t})}=c_{G}-A_{G}\cos(b_{G})\;. \label{coeffmatch2}
\end{equation}
We can solve~(\ref{coeffmatch1}) and~(\ref{coeffmatch2}) simultaneously to find $c_t$ and $a_t$ for the approximate transformation~(\ref{thetaGtan}).
A comparison of the two approximate transformations are given in figure~B\ref{ThetaGandGappfig}.

We can invert~(\ref{thetaGtan}) and find the functional form we require:
\begin{equation}
\phi(\theta)=2\frac{\sqrt{1-a_{t}^{2}}}{1-a_{t}}\tan^{-1}\left(\frac{c_{t}\sqrt{1-a_{t}^{2}}}{\sqrt{2D}}\frac{\theta}{2}\right) \;.
\end{equation}
Finally, we need to find the limits of the domain in the $\theta$ variable. Substituting the $\phi$ limits $\pm \pi$ into~(\ref{thetaGtan}) we find
$\theta \in [-\theta_0,\theta_0]$, with
\begin{equation}
 \theta_0=\frac{\sqrt{2D}}{c_t\sqrt{1-a_t^2}}\;. \label{theta0}
\end{equation}


\section*{References}
\begin{thebibliography}{10}
\item[] \bibitem{Trends} Black A J and McKane A J, {\it Stochastic formulation of ecological models and their applications} 2012
 {\it Trends in Ecology \& Evolution}, {\bf 27} 337
\item[] \bibitem{Murray} Murray J D, 1993 {\it Mathematical Biology},(Berlin:Springer),  2nd corrected ed.
\item[]\bibitem{vanKampen} van Kampen N G, 2007 {\it Stochastic Process in Physics and Chemistry}, (Amsterdam: Elsevier), 3rd ed.
\item[]\bibitem{MckanePredPrey} McKane A J and Newman T J, {\it Predator-Prey Cycles from Resonant Amplification of Demographic Stochasticity}, 2005
 {\it Phys. Rev. Lett.} {\bf 94} 218102
  \item[] \bibitem{Oliveira} Tom\'e T and de Oliveira M J, {\it Role of Noise in Population Dynamics Cycles}, {\it Phys. Rev. E.} {\bf 79} 061128
\item[] \bibitem{StoAmpBioChem} McKane A J \etal, {\it Amplified biochemical oscillations in cellular systems}, 2007 {\it J. Stat. Phys.} {\bf 128}  165
\item[] \bibitem{StoAmpEpi} McKane A J and Black A J, {\it Stochastic amplification in an epidemic model with seasonal forcing}, 2010 {\it J. Theor.
Biol.} {\bf 267} 85
\item[] \bibitem{MobGeoTau} Mobilia M, Georgiev I T and T\"auber U, {\it Phase Transitions and Spatio-Temporal Fluctuations in Stochastic Lattice Lotka-Volterra Models},
 2007 {\it J. Stat. Phys.} {\bf 128}  447
\item[] \bibitem{ParkerE} Parker M and Kamenev A, {\it Extinction in the Lotka-Volterra model}, 2009 {\it Phys. Rev. E.} {\bf 80} 021129
\item[] \bibitem{ParkerJ} Parker M and Kamenev A, {\it Mean Extinction Time in Predator-Prey Model}, 2010 {\it J. Stat. Phys.} {\bf 141} 201
\item[] \bibitem{Abta} Abta R, Schiffer M, Avishag B-I and Shnerb N M, {\it Stabilization of Metapopulation Cycles:Toward a Classification Scheme}, 2008 
{\it Theor. Poul. Biol.} {\bf 74} 273
\item[] \bibitem{Stenseth} Stenseth N C, {\it Population Cycles in Voles and Lemmings: Density Dependence and Phase Dependence in a Stochastic World},
 1999 {\it Oikos} {\bf 87}  427
\item[]\bibitem{Optic} Roichman Y \etal, {\it Influence of Nonconservative Optical Forces on the Dynamics of Optically
	Trapped Colloidal Spheres: The Fountain of Probability}, 2008 {\it Phys. Rev. Lett.} {\bf 101} 128301
 \item[] \bibitem{Risken} Risken H, 1996 {\it The Fokker-Planck Equation: Methods of Solutions and Applications},
 Springer Series in synergetics (Springer), 2nd ed.
  \item[] \bibitem{McKanePatches} McKane A J and Newman T J, {\it Stochastic models in population biology and their deterministic analogs}, 2004
 {\it Phys. Rev. E} {\bf 70} 041902
\item[]\bibitem{Gardiner} Gardiner C, 2009 {\it Stochastic Methods: A Handbook for the Natural and Social Sciences}, (Berlin: Springer), 4th ed.
\item[] \bibitem{Gillespie} Gillespie D T, {\it A general method for numerically simulating the stochastic time evolution
	of coupled chemical reactions}, 1976 {\it J. Comput Phys.} {\bf 22} 403
\item[] \bibitem{ParsonsQuince} Parsons T L and Quince C, {\it Fixation in haploid populations exhibiting density dependence II:
	The quasi-neutral case}, 2007 {\it Theor. Popul. Biol.} {\bf 72} 468
\item[] \bibitem{nullcline} Constable G W A, McKane A J and Rogers T, {\it Stochastic dynamics on slow manifolds} 2013 {\it arXiv:1201.7697v1}
\item[] \bibitem{Kramers} Kramers H A, {\it Brownian motion in a field of force and the diffusion model of chemical reactions}, 1940 {\it Physica} {\bf7} 284
\end{thebibliography}
\end{document}